\documentclass[prb,twocolumn]{revtex4}
\usepackage{graphicx}

\def \nn{\nonumber}
\def \half{ \frac{1}{2} }
\def \nd{{^{\vphantom{\dagger}}}}
\def \yd{^\dagger}
\def \be{\begin{equation}}
\def \ee{\end{equation}}
\def \bra{\langle}
\def \ket{\rangle}
\def \im{ { i } }
\def \siteno{ {N_s} }
\def \siter{ { \mathbf{r} } }
\def \veck{ \mathbf{k} }
\def \pauli{ \sigma }
\def \Tone{ { T_{1} } }
\def \Ttwo{ { T_{2} } }
\def \mirror{ {\sigma} }
\def \rotsix{ {R_{\frac{\pi}{3}}} }
\def \rottwo{ {R_{\pi}} }
\def \PSG#1{ { G_{#1} } }
\def \Gphase#1{ { \phi_{#1} } }
\def \IGGid{ {\rm \bf 1} }
\def \IGGgen{ -{\rm \bf 1} }
\def \SGid{ {\mathcal I} }
\def \idmat{ {\rm \bf 1} }
\def \zeroindex{ ^{[\mathrm{zero-flux}]} }
\def \piindex{ ^{[\pi\mathrm{-flux}]} }
\def \sublatone{ u }
\def \sublattwo{ v }
\def \sublatthree{ w }
\def \spinon{ \Psi }
\def \dif {\Delta}
\def \quo {"}

\usepackage{amsmath}
\begin{document}
\title{Spin Liquid States on the Triangular and Kagom\'e Lattices: A Projective Symmetry Group Analysis of Schwinger Boson States}
\author{Fa Wang$^1$ and Ashvin Vishwanath$^{1,2}$}
\affiliation{$^1$Department of Physics, University of California, Berkeley, CA 94720\\
$^2$Materials Sciences Division, Lawrence Berkeley National
Laboratory, Berkeley, CA 94720\\}
\date{Printed \today}

\begin{abstract}
A symmetry based analysis (Projective Symmetry Group) is used to
study spin liquid phases on the triangular and Kagom\'e lattices in
the Schwinger boson framework. A maximum of eight distinct $Z_2$
spin liquid states are found for each lattice, which preserve all
symmetries. Out of these only a few have nonvanishing nearest
neighbor amplitudes which are studied in greater detail. On the
triangular lattice, only two such states are present - the first
(zero-flux state) is the well known state introduced by Sachdev,
which on condensation of spinons leads to the 120 degree ordered
state. The other solution which we call the $\pi$-flux state has not
previously been discussed. 
Spinon condensation leads to an ordering wavevector at the Brillouin
zone edge centers, in contrast to the 120 degree state. While the
zero-flux state is more stable with just nearest-neighbor exchange,
we find that the introduction of either next-neighbor
antiferromagnetic exchange or four spin ring-exchange (of the sign
obtained from a Hubbard model) tends to favor the $\pi$-flux state.
On the Kagom\' e lattice four solutions are obtained - two have been
previously discussed by Sachdev, which on spinon condensation give
rise to the $q=0$ and $\sqrt{3}\times\sqrt{3}$ spin ordered states.
In addition we find two states with significantly larger values
of the quantum parameter at which magnetic ordering occurs. For one
of them this even exceeds unity, $\kappa_c\approx 2.0$ in a nearest
neighbor model, indicating that if stabilized, could remain spin
disordered for physical values of the spin. This state is also
stabilized by ring exchange interactions with signs as derived from
the Hubbard model.
\end{abstract}
\maketitle

\section{Introduction}\label{sec:motivation}
Recent experiments on frustrated quantum magnets with unusual properties
\cite{Coldea:CsCuCl,Kanoda,He3,Helton-YSLee:kagomespinhalf} has revived interest in spin liquid
phases. At the same time the theoretical understanding of spin
liquid phases in $D>1$ has also matured over the last few years -
for example it is now established that such states are typically
accompanied by emergent gauge fields in their deconfined phase.
Simple and tractable lattice Hamiltonians that realize these phases
can also be constructed. Nevertheless, there is still a pressing
need to understand the situations in which microscopic spin models
might exhibit such phases, and the variety of such phases and their
phenomenological properties.

The triangular lattice antiferromagnet has been considered a good
place to look for such spin liquid states, its geometry is
considered less conducive to magnetic order than for example the
square lattice antiferromagnet, and more recently there has been
evidence from numerical calculations
\cite{Misguich:ExactDiagTrigLattMultipleSpinPhaseDiagram, Sorella:,
Sheng:,Singh:} especially in the presence of anisotropic
interactions and ring exchange, for unconventional physics.
Similarly, recent experiments on quantum magnets with a triangular
lattice structure - Cs$_2$CuCl$_4$ \cite{Coldea:CsCuCl},
$\kappa$-(ET)$_2$Cu(CN)$_3$\cite{Kanoda} and $^3$He films adsorbed on
graphite\cite{He3} show unusual properties that may be interpreted
in terms of spin liquid phases or proximity to such a state.  While
the spin gapped $Z_2$ spin liquid states considered in this paper
are probably not appropriate to directly describe these experimental
systems, except perhaps at the very lowest temperatures, they
motivate a closer study of possible spin liquid states on the
triangular lattice. The highly frustrated Kagom\'e lattice has long
been considered a good system for realizing exotic ground states.
While there are fewer experimentally well studied examples of
quantum spins on the Kagom\'e lattice, conventional semiclassical
analysis fails to produce a unique ground state when applied to this
frustrated lattice. Hence, spin disordered states like spin liquids
may be realized.

In describing spin liquid states of quantum magnets, two distinct
approaches have been found to be useful. The first involves writing
the quantum spin (e.g. spin-1/2) as a fermion bilinear (fermionic
spinons), and imposing a constraint on the total number of fermions
on each site. This fermionic approach is analytically pursued by
writing down a mean field ansatz where the constraint is imposed
only on average to begin with.  This mean field theory may also be
viewed as a source of variational wave-functions, which after
projection yields spin wave functions. An important step in
understanding the variety of spin liquid states possible in this
representation was taken by Wen and collaborators\cite{WenXG:QuantumOrder,
WenXG:SymmetricSpinLiquid, ZhouY:QuantumOrderCsCuCl} who introduced the Projective Symmetry
Group (PSG) classification of spin liquids that respect the
microscopic symmetries of the system. This provides a guide to
identifying the correct gauge group describing the spin liquid (if
it survives projection) and also allows for  a further
classification of different spin liquid phases with the same gauge
group. These different spin liquid phases differ in subtle ways from
one another, for example in terms of the location of the wavevector
of their lowest energy spin-1 excitation. The advantage of such 
classification is that it allows for an exploration of possible spin
liquid states independent of specific Hamiltonians that might
realize them as ground states. In this way the symmetric spin liquid
states within the fermionic spin representation on the square and
triangular lattices were classified \cite{WenXG:SymmetricSpinLiquid,
ZhouY:QuantumOrderCsCuCl}.

A different way to describe spin liquids is via the Schwinger boson
approach, where the spin operator is written as a boson bilinear 
(bosonic spinons) and was analyzed using a large-N approach
\cite{Avoras:SBLargeN, Read:SpNRectLatt, Sachdev:RectLatt,
Sachdev:KagomeAndTrigLatt} . An advantage of this approach is that
it is able to readily access both spin liquid states as well as
conventional magnetically ordered states, which arise from the
condensation of the Schwinger bosons. For instance, in the case of
the triangular lattice\cite{Sachdev:KagomeAndTrigLatt}, Sachdev
found a spin liquid state which on decreasing the strength of
quantum fluctuations to the size expected for s=1/2, gave way to the
120 degree magnetically ordered state, expected of classical spins
on this lattice. However the possibility of other Schwinger boson
spin liquid states on the triangular lattice have not been
systematically explored. Similarly, on the Kagom\'e lattice, Sachdev
\cite{Sachdev:KagomeAndTrigLatt} found two solutions, which on
spinon condensation gives rise to $\sqrt{3}\times\sqrt{3}$ or the
$q=0$ magnetically ordered states.

Mapping out the possible spin liquid states on these lattices is of
interest for two reasons. First, when attempting to identify a
candidate spin liquid state to match with experimental data or
numerical simulations, one needs to specify precisely what the
nature of the state is, and the quantum numbers of its excitations.
Therefore while the simplest spin liquid may fail to match these
details, other spin liquid states may give different predictions.
Second, there may exist solutions stabilized with different
interactions which remain spin liquids even for physical values of
the spin. This would be particularly appealing as it would also
provide direction to the search for spin liquid states. Even when
the system is in the ordered phase, if it is proximate to a critical
point where spin liquid physics holds, it has been argued
\cite{IsakovSenthilKim} 
that finite energy and finite temperature signatures of this will be
unusual and best described in terms of the spin liquid variables.
Therefore, a systematic investigation of the possible solutions of
the Schwinger bosons states is required on the lines of the PSG
analysis of the fermionic spin liquid states. This will also shed
light on the connection between two approaches, for example, which
states can be described in one but not the other approach.

In this paper we will adapt the technique of Projective Symmetry
Groups, developed by Wen and collaborators\cite{WenXG:QuantumOrder,
WenXG:SymmetricSpinLiquid, ZhouY:QuantumOrderCsCuCl} in the context
of fermionic mean field theories to study Schwinger boson mean field
theories. We will focus on the triangular and Kagom\'e lattices, in
particular on spin liquid states with the Ising gauge group ($Z_2$
gauge theories) which preserve the microscopic spin symmetries.
Surprisingly, the number is not large (less than or equal to 8 for
both lattices), which includes, in addition to the solution found by
Sachdev, a set of fundamentally different mean field wave functions, which
give rise to $Z_2$ spin liquids.
In particular, if we make the further reasonable assumption of
restricting to states with non-trivial nearest-neighbor bond
amplitudes, there is only one other state on the triangular lattice,
which we call the $\pi$-flux state. Similarly, on the Kagom\'e
lattice, $Z_2$ spin liquid states  with nonvanishing nearest
neighbor bond amplitudes, include, in addition to the two states
found by Sachdev, two additional states, which we refer to as the
[$\pi$ Hex,$\pi$ Rhom] and [$0$ Hex,$\pi$ Rhom] (which specifies the
flux in the length-six hexagonal loop, and the length-eight
rhombus). One of the problems in searching for spin liquid states
has been the tendency of various spin Hamiltonians to yield
magnetically order states, which is reflected in the small values of
$\kappa_c$, the critical quantum parameter at which spin ordering
occurs in the Sachdev states: $\kappa_c \approx 0.3$ on the
triangular, and $\kappa_c \approx 0.5$ for the Kagom\'e states in
large-N\cite{Sachdev:KagomeAndTrigLatt}. (Note, $\kappa=2S$ in the
case of spin $S$ but it is convenient to extend it to take on all
real values). A feature of the new states that is their relative
stability against magnetic order which is seen in the larger
critical quantum parameter at which spin ordering occurs: $\kappa_c
\approx 0.75$ for the pi-flux triangular state, and $\kappa_c
\approx 0.9$ and $\kappa_c \approx 2.0$ for the two new Kagom\'e
states respectively, in nearest neighbor models.

While the distinct mean field solutions of the PSG analysis are
typically local minima of the mean field energy, an important
question is whether they can be stabilized as global minima with
appropriate interactions. This is discussed in detail for the case
of the triangular lattice in this paper. While the zero-flux state
is stabilized with just nearest-neighbor antiferromagnetic spin
couplings, introducing next neighbor antiferromagnetic couplings or
ring exchange (of the sign arising from the Hubbard model) both tend
to favor the $\pi$-flux state. We obtain the conditions on the
microscopic spin interactions which favor this state and argue that
this may be a fruitful parameter regime for numerical searches for
spin liquids. In general we find that ring exchange on even length
loops (with the sign as derived from the Hubbard model), favors
states with $\pi$ flux on length-$4n$ loops, while it
favors states with zero flux in length-($4n+2$) loops ($n$ is positive integer). This is
in contrast to fermionic case where ring exchange {\em disfavors}
flux\cite{Motrunich:ProjectedFermiSea} for fermions in all loops.
Addition or ring exchange on length-$6$ and length-$8$ loops can stabilize
the [$0$ Hex,$\pi$ Rhom] state on the Kagom\'e, which is good spin
liquid candidate since it has $\kappa_c=2.0$.

 On spinon condensation, the triangular lattice $\pi$-flux state naturally
leads to magnetically ordered states with wavevectors at the
midpoints of the Brillouin zone edges (in contrast to the zero-flux
state which leads to the 120 degree state with wavevector at the
zone corners). This allows us to understand semi-classical (large
spin) calculations in the presence of moderate-neighbor
antiferromagnetic couplings or \quo antiferromagnetic\quo, (with sign as
obtained from the Hubbard model) ring exchange on the triangular
lattice, where ordered states with the same wavevector are found. It
is tempting to connect the possible ordered states on the triangular
lattice as emerging from spin condensation out of the few Schwinger
boson spin liquid states allowed by the PSG. New quantum transitions
are expected on spinon condensation out of the spin liquid states
obtained here, and will be the subject of future study.

{\em Layout of the paper:} In Section~\ref{sec:SBMFT} we briefly
review the Schwinger boson mean field theory. Section~\ref{sec:PSG}
analyzes possible spin liquid states on the triangular lattice. It
first reviews the Projective Symmetry Group classification of spin
liquid states and the strong constraints that arise from relations
between symmetry group elements. This is then applied to Schwinger
boson states on the triangular lattice. A new state is found, the
$\pi$-flux state, which is further analyzed - in particular spin
configurations resulting from spinon condensation are described, and
Hamiltonians stabilizing this mean field solution are obtained. The
general effect of ring exchange interactions on Schwinger boson mean
field states is discussed. In Section \ref{sec:KagomePSG} possible
spin liquid states on the Kagom\'e lattice are studied, and the
properties of one of them, which is unusually stable against spin
ordering, is described in more detail. The PSG analysis and other
details are relegated to the appendices, which also contains
analysis for other lattices of interest such as the anisotropic
triangular lattice.

\section{Schwinger Boson Mean Field Theory}\label{sec:SBMFT}
There are a variety of ways of formulating the Schwinger boson mean
field theory, for example, as a large-N approach
\cite{Avoras:SBLargeN, Read:SpNRectLatt}, or as an approximate
variational approach.

Here we will formulate a variational approach that will provide us
with a unified way to study the effect of different interactions. We
write the spin Hamiltonian:

\be
H=J_{1}\sum_{\bra ij \ket}\mathbf{S}_i\cdot\mathbf{S}_j + J_{2}\sum_{\bra
\bra ij \ket \ket}\mathbf{S}_i\cdot\mathbf{S}_j+\cdots \label{equ:Spin_H}
\ee

in terms of Schwinger bosons:

\be \mathbf{S}_i\cdot\mathbf{S}_j = \frac14 b\yd_{i\,\sigma}
\mathbf{\pauli}_{\sigma,\sigma'}b_{i\,\sigma'}\cdot  b\yd_{j\,\tau}
\mathbf{\pauli}_{\tau,\tau'}b_{j\,\tau'} \label{equ:Spin_Boson} \ee

with the constraint that at every site:

\be \sum_\sigma b\yd_{i\,\sigma}b_{i\,\sigma} = \kappa \label{equ:hardConstraint}\ee

where for a spin system with spin $S$, $\kappa=2S$. In the analysis
below, it will be convenient to consider $\kappa$ to be a continuous
parameter, taking on any non-negative value.

We now consider a variational approach to finding the ground states
and excitations of (\ref{equ:Spin_H}). Motivated by the operator identity
\be
\mathbf{S}_i\cdot\mathbf{S}_j=:\hat{B}_{ij}\yd \hat{B}_{ij}\nd:-\hat{A}_{ij}\yd \hat{A}_{ij}\nd
\ee
where $:\,:$ is normal ordering, and operators $\hat{A}$ and $\hat{B}$ are defined as
\begin{eqnarray}
\hat{B}_{ij}&=&\half\sum_{\sigma}b_{i\sigma}\yd b_{j\sigma}\nd\\
\hat{A}_{ij}&=&\half\sum_{\sigma,\sigma'}\epsilon_{\sigma\sigma'}b_{i\sigma}b_{j\sigma'}
\end{eqnarray}
we consider a \quo Mean Field\quo\ Hamiltonian which is quadratic in terms of the Schwinger bosons,
\begin{eqnarray}\nonumber
H_{MF}&=& \sum_{ij}J_{ij}
(-A_{ij}^*\hat{A}_{ij}+B_{ij}^*\hat{B}_{ij} +{\rm h.c.})\\
\nonumber &&+\sum_{ij}{{J_{ij}}(A_{ij}^*A_{ij}-B_{ij}^*B_{ij})} \\
&& - \mu\cdot\sum_{i}{ \left (\sum_{\sigma}{b_{i\sigma}\yd
b_{i\sigma}}\nd - \kappa\right ) } \label{equ:mfHamiltonian}
\end{eqnarray}
where complex numbers $A_{ij}\nd =-A_{ji}\nd ,\,B_{ij}\nd =B_{ji}^*$
are the parameters of the mean field ansatz.

In the large-N $Sp(N)$ theory, the mean field Hamiltonian contains
only the large-N generalization of the $A$ term. However, since both
$A$ and $B$ terms are consistent with global $SU(2)$ symmetry
(global spin rotation symmetry) they are both included in the
current theory. The introduction of both terms can be found in Gazza
\cite{Gazza:SBBothAB} and many other papers
\cite{Lefmann:SBandNeutronScattering,
Misguich:SBMultipleSpinPhaseDiagram}.

This Hamiltonian is used to
generate a variational wavefunction in terms of the variational
parameters $|\Psi(A_{ij},\,B_{ij},\mu)\ket$. In order to obtain a
{\it spin} wavefunction, we need to project $|\Psi\ket$ into the
constrained Hilbert space where the total number of bosons at each
site is exactly $2S$. Strictly speaking one must evaluate
variational energies after this projection step, using the spin
Hamiltonian (\ref{equ:Spin_H}). This generalization of Gutzwiller projection 
to Schwinger boson has been studied by Chen and collaborators\cite{ChenYC:trigAFMprojection,ChenYC:squareAFMprojection}. 
However, since this hard projection is
not possible to implement analytically (it is even difficult to do
numerically) we rely on a approximate strategy that forgoes
implementing the constraint locally, but only on the average - i.e.
we tune $\mu$ to ensure that:
\be
\sum_{\sigma}{ \bra b\yd_{i\,\sigma}b_{i\,\sigma}\ket }= \kappa \label{equ:avgConstraint}
\ee
we then evaluate the expectation value of the Hamiltonian
(\ref{equ:Spin_Boson}), written out in terms of Schwinger bosons
(\ref{equ:Spin_Boson}) using the variational wavefunction. The
resulting variational energy is then minimized with respect to the
variational parameters $A_{ij},\ B_{ij}$. This yields the
self-consistent equations, 
\be
\bra\hat{A}_{ij}\ket=A_{ij},\,\quad\bra\hat{B}_{ij}\ket=B_{ij} 
\ee

When these self-consistent equations are satisfied,
the variational(mean field) energy is simply obtained by utilizing the following identity,
\be
\bra\mathbf{S}_i\cdot\mathbf{S}_j \ket = \frac32\left ( \vert
B_{ij}\vert^2 - \vert A_{ij}\vert^2 \right )\label{equ:SdotS}
\ee
The effect of projection on energetics will be a topic of future study.

\section{PSG for Schwinger Boson States on the Triangular Lattice}\label{sec:PSG}
We would like to classify Schwinger boson mean field states
available to us on the triangular lattice in terms of the distinct
spin liquid phases they can give rise to. This exercise allows us to
readily obtain potentially interesting states relevant to quantum
spin systems on the triangular lattice. Interactions that stabilize
these states are discussed subsequently.

This classification follows the  classification of fermionic mean
field states in \cite{WenXG:QuantumOrder, WenXG:SymmetricSpinLiquid,
ZhouY:QuantumOrderCsCuCl}. Here however the underlying local
transformation is $U(1)$ (not $SU(2)$ as in the case of fermions).
For mean field theories that survive projection (or equivalently,
survive fluctuations) to give rise to a spin liquid state, this
procedure classifies the distinct quantum phases of the system. We
will require that the mean field theory reflects the underlying
microscopic symmetries of the spin model. This leads to symmetric
spin liquid states. The symmetry transformations include the space
group (lattice translations and point group symmetries) of the
triangular lattice, spin rotation symmetry and time reversal
symmetry. The novel ingredient here is that the mean field state may
preserve these symmetries in some indirect way. In addition to the
symmetries of the original spin model, one also finds an extra
global symmetry in all the mean field ansatz which is a subset of
the local $U(1)$ transformations mentioned previously. This is
called the Invariant Gauge Group (IGG) in the terminology of
Ref.~\cite{WenXG:QuantumOrder}. Since we are interested in mean
field states that have exactly the same symmetry as the underlying
spin model, not less or more, this is to be identified with the
gauge group of the emergent gauge theory. In the examples below we
will find an extra $Z_2$ symmetry in the mean field ansatz, hence
the spin liquids that are obtained with this procedure are $Z_2$
spin liquids, but with different internal structures depending on
how the microscopic symmetries of the lattice model are realized.
Thus, in contrast to conventional states which are distinguished by
patterns of broken symmetry, spin liquid phases that are completely
symmetric and even share the same gauge group can be further
distinguished in terms of how the symmetries are realized.

In Schwinger boson representation of spins, 
there is a local $U(1)$ transformation of bosons:
\be b_{\siter\sigma}\rightarrow
e^{\im\Gphase{}(\siter)}b_{\siter\sigma} \label{equ:localU1} \ee
which leaves all physical observables unchanged. Under this
transformation, our mean field ansatz ($A_{ij}$, $B_{ij}$) also
transform:
\begin{subequations}
\begin{eqnarray}
A_{ij}&\rightarrow&e^{-\im\Gphase{}(i)-\im\Gphase{}(j)}A_{ij}
\label{equ:localU1A}\\
B_{ij}&\rightarrow&e^{+\im\Gphase{}(i)-\im\Gphase{}(j)}B_{ij}
\label{equ:localU1B}
\end{eqnarray}
\end{subequations}
Two mean field ansatz that are related by such a transformation,
give rise to the same spin wave function after projection. However,
we will be interested in a class of transformations that leave the
mean field ansatz {\em invariant}. Naively, one might expect that
since we are interested in states that maintain all the microscopic
symmetries, the mean field ansatz should be invariant under their
operation ({\it e.g.} lattice translations). While this is a sufficient
condition for invariance, it is not a necessary one, due to the
presence of the local $U(1)$ transformations described above. A
symmetry operation might return the ansatz to a $U(1)$ transformed
form which would suffice, since the same spin wavefunction will be
obtained on projection. Hence symmetry transformations that leave
the Ansatz invariant in general will contain the naive
transformation combined with a local $U(1)$ transformation. The set
of all transformations that leave a mean ansatz invariant is called
the {\em Projective Symmetry Group (PSG)} \cite{WenXG:QuantumOrder}.
In principle, we would like to associate each and every element of
this group with a physical symmetry. This is because if the mean
field ansatz is to be a faithful representation of the microscopic
physics, it should have exactly as much symmetry as the original
model. However, we always find that there are some elements of the
PSG that are pure local transformation of the kind
(\ref{equ:localU1}). The set of such elements also forms a group (a
subgroup of the PSG) and is called the invariant gauge group (IGG).
These cannot be the result of a physical symmetry. It is therefore
natural to associate these elements with the emergent gauge group
that describes the spin liquid phase obtained (if the mean field
state survives projection). Therefore the first step is to identify
the IGG of a mean field ansatz.

{\em The Invariant Gauge Group for Schwinger Boson Mean Field
Theories:} A general Schwinger boson mean field Hamiltonian with
explicit global $SU(2)$ symmetry (global spin rotation symmetry)
must be of the form of equation (\ref{equ:mfHamiltonian}).

It is clear that if $A_{ij}$ and $B_{ij}$ are both nonzero the IGG
must be a $Z_2$ group. The only two elements of IGG are identity
operation $\IGGid$ and the IGG generator $\IGGgen:\ b_{\siter
\sigma}\rightarrow -b_{\siter\sigma}$. This can be seen by
considering only one bond.

Note, that if $B_{ij}$ are nonzero while all $A_{ij}$ vanish, the
IGG will be a $U(1)$ group: $b_{\siter \sigma}\rightarrow
e^{\im\Gphase{}}b_{\siter\sigma}$, where $\Gphase{}$ is a
site-independent constant. However it is unlikely for this ansatz to
describe an antiferromagnet according to (\ref{equ:SdotS}). Finally
we should consider the case with only non-vanishing $A_{ij}$. On a
frustrated lattice the IGG will still be the above $Z_2$ group.
However, if the lattice is bipartite, the IGG will be a $U(1)$
group: $b_{\siter\sigma}\rightarrow
e^{\pm\im\Gphase{}}b_{\siter\sigma}$, where we apply opposite signs
on the two sublattices. This is the case of simple square lattice.

We will only consider PSGs with $Z_2$ IGG in the following
discussion. Hence we are implicitly restricting ourselves to $Z_2$
spin liquid states, which are the natural spin liquid states on
frustrated lattices within the Schwinger boson formalism.

\subsection{Algebraic Constraints on the PSG}
We consider mean field Hamiltonian that preserves all of the
physical symmetries in the PSG sense (symmetric spin liquid states).
They are spin rotation symmetry, lattice space group symmetries and
time reversal symmetry. The spin rotation symmetry is already
implemented by considering mean field ansatz of the form
(\ref{equ:mfHamiltonian}), which is explicitly invariant under
global $SU(2)$ spin rotations. We will consider time reversal
symmetry at the end of our derivation. The operations that we will
now pay special attention to are the space group symmetries
(translations and point group operations for the triangular
lattice). As discussed before, these can be implemented via
combining the naive transformation with a local (gauge) $U(1)$
transformation. One can ask the question - is it possible to have
mean field ansatz with any choice of the gauge transformations? It
turns out that there are algebraic relations among the symmetry
group elements that strongly constrain the possible choices of gauge
transformations.
Thus, to obtain all possible PSGs we should first check the
algebraic structure of PSGs without reference to a specific mean
field ansatz. The possible PSGs allowed by the algebra of the space
group are defined as the {\em algebraic PSGs} related to the space
group.

For the isotropic triangular lattice, the space group is generated
by two translations $\Tone$ and $\Ttwo$, one reflection $\mirror$ in
a bond and the 60 degree rotation $\rotsix$ about a lattice site.

We use the following oblique coordinate system. In this system a
site index $\siter$ has two integer components $\siter=(r_1,r_2)$.
\begin{figure}[h]
\includegraphics{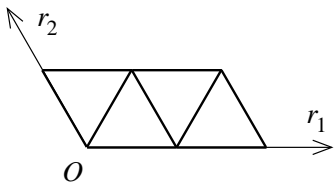}
\caption{\label{fig:coordinatessystem}Oblique coordinates system of triangular lattice.}
\end{figure}
Then the four generators are given by the following formulae
\begin{subequations}
\begin{eqnarray}
\Tone:\ (r_1,\ r_2)&\rightarrow&(r_1+1,\ r_2)\\
\Ttwo:\ (r_1,\ r_2)&\rightarrow&(r_1,\ r_2+1)\\
\mirror:\ (r_1,\ r_2)&\rightarrow&(r_2,\ r_1)\\
\rotsix:\ (r_1,\ r_2)&\rightarrow&(r_1-r_2,\ r_1)
\end{eqnarray}
\end{subequations}

For symmetric spin liquid states there must be four gauge group
transformations $\PSG{\Tone}$, $\PSG{\Ttwo}$, $\PSG{\mirror}$, and
$\PSG{\rotsix}$, such that the mean field ansatz is invariant under
$\PSG{\Tone}\Tone$, $\PSG{\Ttwo}\Ttwo$, $\PSG{\mirror}\mirror$, and
$\PSG{\rotsix}\rotsix$, respectively. We can represent the four
gauge operations by their phase, \be \PSG{X}:\
b_{\siter\sigma}\rightarrow
e^{\im\Gphase{X}(\siter)}b_{\siter\sigma} \ee where the $X$ are
$\Tone$, $\Ttwo$, $\mirror$, and $\rotsix$, respectively.

The PSG is generated by combining the generators of the IGG:
$\IGGgen$, and the above four compound operators $\PSG{X}X$.

\begin{figure}[h]
\includegraphics{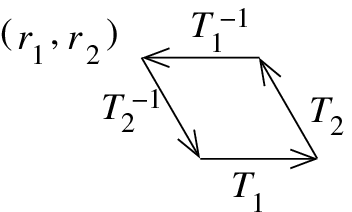}
\caption{\label{fig:algebraicrelation}An example of a relation
between space group elements $\Tone^{-1}\Ttwo\Tone\Ttwo^{-1}=\SGid$,
that leads to an algebraic constraint on the PSG.}
\end{figure}

The structure of the space group imposes algebraic constraints on
the $\PSG{X}$. For instance, the combination of translations shown
in FIG.~\ref{fig:algebraicrelation} should be equivalent to the
identity, i.e.  $\Tone^{-1}\Ttwo\Tone\Ttwo^{-1}=\SGid$. Therefore we
require that the implementation of these symmetries in the PSG:
$(\PSG{\Tone}\Tone)^{-1}\PSG{\Ttwo}\Ttwo\PSG{\Tone}\Tone(\PSG{\Ttwo}\Ttwo)^{-1}$
must be the equivalent of an identity operation, which means it is
an element of the IGG, either $\IGGid$ or $\IGGgen$. This string can
be rewritten as $\Tone^{-1}\PSG{\Tone}^{-1}\Tone \cdot
\Tone^{-1}\PSG{\Ttwo}\Tone \cdot
(\Tone^{-1}\Ttwo)\PSG{\Tone}(\Tone^{-1}\Ttwo)^{-1} \cdot
(\PSG{\Ttwo})^{-1}$. Using the fact that for a space group operation
$Y$, the gauge transformation $(Y)^{-1}\cdot \PSG{X}\cdot Y$ acting
on site $\siter$ will just give a phase $\Gphase{X}[Y(\siter)]$
(where $Y(r)$ is the image of $r$ under the space group operation
$Y$), we end up with the equation 
\be
\begin{split}
&-\Gphase{\Tone}[\Tone(\siter )]+\Gphase{\Ttwo}[\Tone(\siter )]+\Gphase{\Tone}[\Ttwo^{-1}\Tone(\siter )]-\Gphase{\Ttwo}(\siter)\label{equ:PSGT1T2equation} \\
=&p_1\pi
\end{split}
\ee 
where $p_1\in\{0,1\}$ is independent
of the site index $\siter$, and arises from the fact that the
identity operation can be any one of the two elements of the IGG.

This kind of algebraic constraints strictly restrict the possible
structures of PSG. It turns out (proof is in Appendix~\ref{asec:algPSG}) 
that there is only a finite set of such
constraints, which if satisfied guarantees that all relations
between space group elements are satisfied. These relations,
beginning with the one described above, are:
\begin{subequations}
\begin{eqnarray}
\label{eqn:TT}
\Ttwo\Tone &=& \Tone\Ttwo\\
\Tone\rotsix &=& \rotsix \Ttwo^{-1}\\
\Ttwo \rotsix &=& \rotsix \Tone \Ttwo\\
\Tone \mirror &=& \mirror\Ttwo \\
\Tone \mirror &=& \mirror \Ttwo \\
\rotsix^{-1} &=& [\rotsix ]^5 \\
\mirror^{-1} &=&\mirror\\
 \rotsix\mirror &=& \mirror [\rotsix ]^{5}
\label{eqn:mirrorrotsix}
\end{eqnarray}
\end{subequations}

These relations place severe constraints on the allowed PSGs. The
most general solution that satisfies them is:
\begin{subequations}
\begin{eqnarray}
\Gphase{\Tone}(r_1,\ r_2)&=&0\label{equ:PSGphiT1}\\
\Gphase{\Ttwo}(r_1,\ r_2)&=&p_1\pi r_1\label{equ:PSGphiT2}\\
\Gphase{\mirror}(r_1,\ r_2)&=&p_2\pi/2+p_1\pi r_1 r_2 \label{equ:PSGphimirror}\\
2 \Gphase{\rotsix}(r_1,\ r_2)&=&p_3\pi+p_1\pi r_2(r_2-1+2r_1)\label{equ:PSGphirotsix}
\end{eqnarray}
\end{subequations}
where $p_1,\ p_2,\ p_3$ are either 0 or 1. Thus there are at most
{\em eight} distinct Schwinger boson symmetric spin liquid states on
isotropic triangular lattice. If we put more conditions on the mean
field ansatz ({\it e.g.} that the nearest-neighbor $A_{ij}$ are
nonzero), then the number of possible symmetric spin liquid states
is reduced from eight.

A detailed derivation of the above formulae is given in
Appendix~\ref{asec:algPSG}. Also we include the solution of
algebraic PSGs on the anisotropic triangular lattice.

\subsection{From PSGs to Mean Field Hamiltonians :\\ Nearest-neighbor Models}\label{ssec:realization}
If we assume that nearest-neighbor amplitudes $A_{ij}$ are nonzero,
(which is natural given that nearest neighbor interactions in
physical models tend to be dominant and usually antiferromagnetic),
there are more constraints on the possible PSG structures.

This can be seen as follows. Since $\mirror$ maps bond
$(0,0)\rightarrow (1,1)$ to itself, we must have
$\Gphase{\mirror}(0,0)+\Gphase{\mirror}(1,1)=0\mod 2\pi$. This
imposes the constraint $p_2=p_1\mod 2$. Also, bonds
$(0,0)\rightarrow (-1,0)$ and $(0,0)\rightarrow (1,0)$ are related
by a 180 degree rotation, $(\rotsix)^3$, and by antisymmetry  and
translation $\Tone$ we have
$A_{(0,0)\rightarrow (-1,0)}=-A_{(-1,0)\rightarrow (0,0)}=-A_{(0,0)\rightarrow (1,0)}$. This leads to
another constraint \be
\begin{split}
&\Gphase{\rotsix}(-1,0)+\Gphase{\rotsix}(0,1)+\Gphase{\rotsix}(1,1)+3\Gphase{\rotsix}(0,0)\\
=&\pi\mod 2\pi\nn
\end{split}
\nn \ee which fixes $p_3=1-p_1\mod 2$.

Thus we have only {\em two} non-equivalent PSGs, corresponding to
$p_1=0$, which we call the zero-flux state  or $p_1=1$ which we call
the $\pi$-flux state.

{\em 1. The zero-flux state:}

This state is specified in terms of the integers $p_1=0,\ p_2=0,\
p_3=1$, or equivalently in terms of the phase factors involved in
implementing the space group operations:
\begin{eqnarray*}
\Gphase{\Tone}(r_1,\ r_2)=\Gphase{\Ttwo}(r_1,\ r_2)=\Gphase{\mirror}(r_1,\ r_2)&=&0\\
\Gphase{\rotsix}(r_1,\ r_2)&=&\pi/2
\end{eqnarray*}

\begin{figure}[h]
\includegraphics{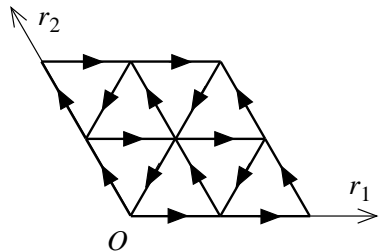}
\caption{\label{fig:zeroansatz} Ansatz of the zero-flux state. An
arrow from site $i$ to site $j$ means $A_{ij}>0$. All $A_{ij}$ have
the same magnitude. All $B_{ij}$ are real and of uniform magnitude
(not shown).}
\end{figure}

We can now ask what mean field ansatz would be characterized by such
a PSG. The mean field ansatz is specified by the amplitudes $A_{ij}$
and $B_{ij}$ on the various bonds. For this PSG, $B_{ij}$ and
$A_{ij}$ on nearest-neighbor are both real and nonzero in general.
Explicit expression for this ansatz are:
\begin{eqnarray*}
&&A_{(r_1,r_2)\rightarrow (r_1+1,r_2)}=A_{(r_1,r_2)\rightarrow (r_1,r_2+1)}\\
&=&-A_{(r_1,r_2)\rightarrow (r_1+1,r_2+1)}=A_1,\\
&&B_{(r_1,r_2)\rightarrow (r_1+1,r_2)}=B_{(r_1,r_2)\rightarrow (r_1,r_2+1)}\\
&=&B_{(r_1,r_2)\rightarrow (r_1+1,r_2+1)}=B_1
\end{eqnarray*}
The PSG predicts that next-nearest-neighbor $A$ must be zero. Since
$A_{ij}=-A_{ji}$ is real and anti-symmetric, it is natural to
represent it by oriented bonds. FIG.~\ref{fig:zeroansatz} is a
pictorial representation of the ansatz for the zero-flux state.

This ansatz has explicit translational invariance, and has been
studied by Sachdev using the large-N method
\cite{Sachdev:KagomeAndTrigLatt}.

{\em 2. $\pi$-flux state:}

This state is specified in terms of the integers
 $p_1=1,\ p_2=1,\ p_3=0$, or equivalently in terms of the phase factors involved in
implementing the space group operations:
\begin{eqnarray*}
\Gphase{\Tone}(r_1,\ r_2)&=&0\\
\Gphase{\Ttwo}(r_1,\ r_2)&=&\pi r_1\\
\Gphase{\mirror}(r_1,\ r_2)&=&(\pi/2)+\pi r_1 r_2\\
\Gphase{\rotsix}(r_1,\ r_2)&=&\pi r_1 r_2+(\pi/2) r_2(r_2-1)
\end{eqnarray*}

\begin{figure}[h]
\includegraphics{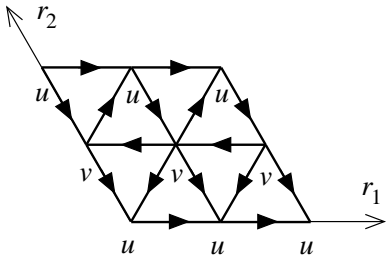}
\caption{\label{fig:piansatz}Ansatz of the $\pi$-flux state. All
nearest-neighbor $A_{ij}$ are real and of the same magnitude, and
are positive in the directions shown. All nearest-neighbor $B_{ij}$
are zero. The unit cell for the Schwinger bosons is doubled due to
the flux, and $\sublatone$ and $\sublattwo$ are the two sites in
this unit cell. }
\end{figure}

The mean field ansatz that realizes this PSG is as follows. While
the nearest-neighbor bond $A_{ij}$ is real and nonzero, the
nearest-neighbor bond $B_{ij}$ must be zero. Expression of the
ansatz is:
\begin{eqnarray*}
&&(-1)^{r_2}A_{(r_1,r_2)\rightarrow (r_1+1,r_2)}=-A_{(r_1,r_2)\rightarrow (r_1,r_2+1)}\\
&=&-(-1)^{r_2}A_{(r_1,r_2)\rightarrow (r_1+1,r_2+1)}=A_1,\\
&&B_{(r_1,r_2)\rightarrow (r_1+1,r_2)}=B_{(r_1,r_2)\rightarrow (r_1,r_2+1)}\\
&=&B_{(r_1,r_2)\rightarrow (r_1+1,r_2+1)}=0
\end{eqnarray*}
The PSG also predicts that the next-nearest-neighbor $B$ must be
zero. Note, for this state the translation symmetry is not explicit
in the Mean Field Hamiltonian, hence the unit cell for the Schwinger
bosons is doubled. This $\pi$-flux ansatz for nearest-neighbor model
is shown in FIG.~\ref{fig:piansatz}.

These two states can also be distinguished by the gauge-invariant
phase of the product $A_{ij}\nd  (-A_{jk}^*) A_{kl}\nd (-A_{li}^*)$,
where $i,j,k,l$ form a rhombus with unit length sides. This quantity
was introduced by Tchernyshyov, {\it et. al.}
\cite{Tchernyshyov:GreedyBoson} as the {\em flux} in the rhombus in
bosonic large-N theory. Defining: \be |A_1|^4e^{i\Phi} = A_{ij}\nd
(-A_{jk}^*) A_{kl}\nd (-A_{li}^*)\label{eqn:flux} \ee where $|A_1|$
is the uniform magnitude of $A_{ij}$, we have that for the zero-flux
state, the flux $\Phi=0$ for all rhombi; while for the $\pi$-flux
state $\Phi=\pi$ for all rhombi. Therefore these two states are
clearly not gauge equivalent mean field states.

Finally we can consider time reversal symmetry $\hat{T}$. Because
the two ansatz are both real, they directly respect
$\hat{T}$-symmetry, since time reversal transformation will change
the ansatz to their complex conjugate. It is then an interesting
question to ask what kind of PSG can support $\hat{T}$-breaking
ansatz. It turns out that one must also break lattice reflection
symmetry to obtain a time reversal breaking state. In
Appendix.~\ref{asec:PSGaniso} we list the solutions of algebraic
PSGs for anisotropic triangular lattice and the realizations in
nearest-neighbor model which do support a $\hat{T}$-breaking ansatz.

\section{Analysis of the Mean Field Theories}
We now study the mean field theories arising from the ansatz
described previously. Given the PSG classification, it follows that
these different mean field solutions will be local minima of the
mean field energy. In order to pick which of these is favored with a
particular Hamiltonian, one needs to compare energies between these
states, which is left to the next section. Here we content ourselves
with describing the properties of each of these mean field ansatz.
The discussion closely parallel Sachdev's large-N solution of
triangular lattice and Kagom\'e lattice
\cite{Sachdev:KagomeAndTrigLatt}. We also take the quantum parameter
$\kappa$ as a continuous number, although for Schwinger bosons
derived from $SU(2)$ spins of size $S$, $\kappa=2S$. For small
$\kappa$, the spinon dispersion will be gapped and we have a spin
liquid mean field ground state. When $\kappa$ goes beyond certain
critical value $\kappa_c$  spinon dispersion becomes gapless,
bosonic spinons condense, and magnetic long range order develops.
For simplicity we present only the solution to the nearest-neighbor
model.

\subsection{Zero-flux state}
The mean field theory of this state is almost identical to Sachdev's
large-N theory for triangular lattice
\cite{Sachdev:KagomeAndTrigLatt} and the phases obtained are
continuously connected to the phases described in that work. The
only difference is that we include the parameter $B_{ij}$, which is
not present in the large-N treatment,and therefore our spinon
dispersion and critical quantum parameter $\kappa_c$ differ
slightly. This is reviewed briefly here before turning to a similar
analysis of the $\pi$-flux state.

\begin{figure}[h]
\includegraphics{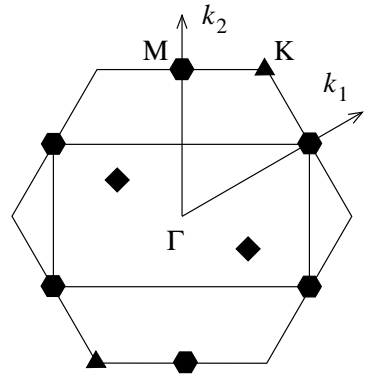}
\caption{\label{fig:BZ} Triangular Lattice Brillouin Zone (BZ) and
$k$-space coordinate system. The large hexagon is the BZ of the
original lattice and the zero-flux ansatz. The large rectangle is
the BZ of the $\pi$-flux ansatz. Black triangles at K-points: 
$\pm(2\pi/3,2\pi/3)$ are the minima of the spinon dispersion of the
zero-flux state, and are also the wavevectors of magnetic ordering
of the zero-flux state (120 degree state). Black diamonds at
$\pm(\pi/2,-\pi/2)$ are the minima of the spinon dispersion of the
$\pi$-flux state. Black hexagons at M-points: 
$\pm(\pi,-\pi),\,\pm(-\pi,0),\,\pm(\pi,0)$ are the magnetic ordering
wavevectors of the $\pi$-flux state. }
\end{figure}

The Brillouin zone for the triangular lattice and our choice of
coordinate systems in  $\veck$-space in shown in FIG.~\ref{fig:BZ}.
In this oblique coordinates system, $\veck\cdot\siter=k_1 r_1+k_2
r_2$. For covenience we define $k_3=-k_1-k_2$.

After Fourier transformation, the nearest-neighbor mean field
Hamiltonian (\ref{equ:mfHamiltonian}) is \be
\begin{split}
H_{MF}=&\phantom{+}\sum_{\veck}{
 \spinon(\veck)\yd
 D(\veck)
 \spinon(\veck)
}\\
&+\siteno[\mu+\mu\kappa-3J_{1}(B_{1}^2-A_{1}^2)]
\end{split}
\nn
\ee

where $\siteno$ is the number of sites and the vector spinon field
$\spinon(\veck)$ and the 2-by-2 matrix $D(\veck)$ are:
\begin{eqnarray}
\spinon(\veck)&=&\begin{pmatrix}b_{\veck\uparrow}\nd\cr
b_{-\veck\downarrow}\yd\end{pmatrix} \\
D(\veck) &=&
\begin{pmatrix}
J_{1} B_{1} \mathrm{Re}(\xi_\veck)-\mu & -\im J_{1}
A_{1}\mathrm{Im}(\xi_\veck)\cr \im J_{1}A_{1} \mathrm{Im}(\xi_\veck)
& J_{1}B_{1}\mathrm{Re}(\xi_\veck)-\mu
\end{pmatrix}
\label{Dzeroflux}
\end{eqnarray}
where $\xi_\veck=e^{\im k_1}+e^{\im k_2}+e^{-\im (k_1+k_2)}$.

After a Bogoliubov transformation
\be
\begin{pmatrix}b_{\veck\uparrow}\nd\cr
b_{-\veck\downarrow}\yd\end{pmatrix}
=M_{\veck}
\begin{pmatrix}\gamma_{\veck\uparrow}\nd\cr
\gamma_{-\veck\downarrow}\yd\end{pmatrix} \ee where $M_{\veck}\in
SU(1,1)$, is chosen to diagonalize the mean field Hamiltonian:
\begin{eqnarray*}
H_{MF}&=&
 \sum_{\veck}{
\omega(\veck)(\gamma\yd_{\veck\uparrow}\gamma\nd_{\veck\uparrow}+
\gamma\yd_{-\veck\downarrow}\gamma\nd_{-\veck\downarrow}+1)
}\\
&& +\siteno[\mu+\mu\kappa-3J_{1}(B_{1}^2-A_{1}^2)]
\end{eqnarray*}
where the spinon dispersion $\omega(\veck)$ is:  \be
\omega(\veck)=\sqrt{\left [J_{1}
B_{1}\mathrm{Re}(\xi_\veck)-\mu\right ]^2 -\left
[J_{1}A_{1}\mathrm{Im}(\xi_\veck)\right ]^2} \ee and the
self-consistency equations are:
\begin{eqnarray}
6J_{1}A_{1}&=&-\int_{BZ}
 \frac{\partial\omega(\veck)}{\partial A_{1}}\mathrm{\bf d}^2k\\
6J_{1}B_{1}&=&+\int_{BZ}
 \frac{\partial\omega(\veck)}{\partial B_{1}}\mathrm{\bf d}^2 k\\
1+\kappa&=&-\int_{BZ} \frac{\partial\omega(\veck)}{\partial
\mu}\mathrm{\bf d}^2k
\end{eqnarray}
where the integral is over the Brillouin Zone, and the integration
measure $\mathrm{\bf d}^2k = {\mathrm d k}_1{\mathrm d k}_2/(4\pi^2)$.

The mean field energy per bond is \be
E/\mathrm{bond}=J_{1}(B_{1}^2-A_{1}^2) \ee

\begin{figure}[h]
\includegraphics{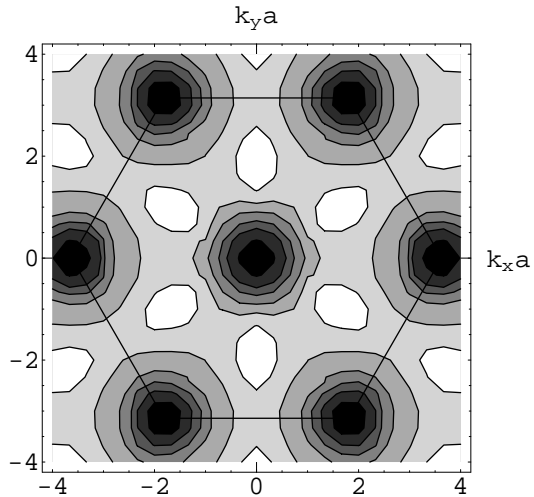}
\caption{\label{fig:zerospinspectrum} The  lower edge of the
two-spinon spectrum for the zero-flux state at $\kappa=0.3$. 
Axes are in dimensionless unit $k_{x,y}a$ where $a$ is lattice constant. 
Darker regions has lower energy. the 120 degree magnetic order arises from
magnon condensation at the zone corners. See also FIG.~\ref{fig:BZ}. }
\end{figure}

Minima of the spinon dispersion are at
$(k_1,\,k_2)=\pm(2\pi/3,\,2\pi/3)$, which are the corners of the
Brillouin zone. The two-spinon spectrum also has minima at the
Brillouin zone corners and at the zone center. As in Ref.
\cite{ZhouY:QuantumOrderCsCuCl}, we display in FIG.
\ref{fig:zerospinspectrum} a contour plot of the minimum energy
required to create a two spinon excitation at a given crystal
momentum (the lower edge of the two-spinon spectrum) to show this
feature.

By solving $\min\omega(\veck)=0$ together with self-consistent
equations, we determined the critical quantum parameter
$\kappa_c\approx 0.42$. We can estimate the sublattice magnetization
for spin-1/2 Heisenberg model to be $58\%$ of classical spin, which
probably overestimates the order since it is larger than that
obtained from spin-wave theories which predicts this to be about
$48\%$ \cite{Jolicoeur:SpinWaveNearestNeighborTrig}.

\subsection{$\pi$-flux state}

From FIG.~\ref{fig:piansatz}, it is clear that there are two sites
in a unit cell, distinguished by the parity of $r_2$. We label them
by $\sublatone$ and $\sublattwo$.

After Fourier transformation, the mean field Hamiltonian
(\ref{equ:mfHamiltonian}) is \be
\begin{split}
H_{MF}=&\phantom{+}\sum_{\veck}{
 \spinon(\veck)\yd
 D(\veck)
 \spinon(\veck)
}\\
&+\siteno[\mu+\mu\kappa+3J_{1}A_{1}^2]
\end{split}
\label{equ:FTedmfHpistate}
\ee
Here $\sum_{\veck}$ sums over $(\siteno/2)$ $\veck$-points in spinon Brillouin zone,
which is one half of the hexagonal Brillouin zone of original spin model (see FIG.~\ref{fig:BZ}).

The vector spinon field $\spinon(\veck)$ is: \be
\spinon(\veck)=\begin{pmatrix} b_{\sublatone\veck\uparrow}\nd\cr
b_{\sublattwo\veck\uparrow}\nd\cr
b_{\sublatone\mathbf{-k}\downarrow}\yd\cr
b_{\sublattwo\mathbf{-k}\downarrow}\yd\cr \end{pmatrix} \ee and the
4-by-4 matrix $D(\veck)$ has the following block form \be
\begin{pmatrix}
\mu\cdot \idmat&-\im P(\veck)\nd\cr
\im P(\veck)\yd&\mu\cdot \idmat
\end{pmatrix}
\label{equ:DPmatrix} \ee where $\idmat$ is the 2-by-2 identity
matrix, and the 2-by-2 Hermitian matrix $P(\veck)$ can be written in
terms of the Pauli matrices $\pauli^{x,y,z}$: \be P(\veck)=J_{1}
A_{1} \left
[\sin(k_1)\pauli^z-\sin(k_2)\pauli^x-\cos(k_3)\pauli^y\right ] \ee

This Hamiltonian can be diagonalized by a Bogoliubov transformation
using a $SU(2,2)$ matrix after which we have:
\begin{eqnarray}\nonumber
H_{MF}&=& \sum_{\veck;\,a=u,v}{ \omega(\veck)(
 \gamma_{a\veck\uparrow}\yd\gamma_{a\veck\uparrow}\nd
+\gamma_{a-\veck\downarrow}\yd\gamma_{a-\veck\downarrow}\nd +1 )
}\\
&& +\siteno[\mu+\mu\kappa+3J_{1}A_{1}^2]
\end{eqnarray}
where $\gamma$ are transformed boson operators and $a=u,v$ is a
sublattice index.

The spinon dispersion is therefore four-fold degenerate with the
form:
\begin{eqnarray*}
\omega(\veck)&=&\sqrt{\mu^2-J_{1}^2 A_{1}^2\left [\sin^2(k_1)+\sin^2(k_2)+\cos^2(k_3)\right ] }
\end{eqnarray*}

Self-consistent equations comes from minimization of mean field energy
(constant terms after Bogoliubov transformation).
\begin{eqnarray}
6J_{1}A_{1}&=&-\int_{BZ}\frac{\partial\omega(\veck)}{\partial A_{1}}\mathrm{\bf d}^2k\\
1+\kappa&=&-\int_{BZ}\frac{\partial\omega(\veck)}{\partial
\mu}\mathrm{\bf d}^2k
\end{eqnarray}
where the integral is over the Brillouin Zone, and $\mathrm{\bf
d}^2k = {\mathrm d k}_1{\mathrm d k}_2/(4\pi^2)$. Recall, the
amplitudes $B_{ij}$ on nearest neighbor bonds are forbidden in this
state by the PSG.

\begin{figure}[h]
\includegraphics{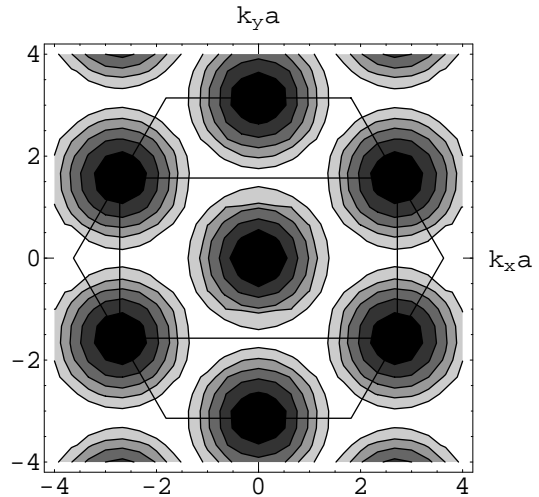}
\caption{\label{fig:pispinspectrum} The lower edge of the two-spinon
spectrum of the $\pi$-flux state with $\kappa=0.3$. 
Axes are in dimensionless unit $k_{x,y}a$ where $a$ is lattice constant. 
Darker regions have lower energy. Magnetic order arises from magnon condensation at 
the centers of the zone edges. See also FIG.~\ref{fig:BZ}.}
\end{figure}

On solving the self-consistent equations, the mean field energy per
bond is obtained as:
 \be E/\mathrm{bond}=-J_{1}A_{1}^2 \ee
and the minima of spinon dispersion are at
$(k_1,k_2)=\pm(\pi/2,-\pi/2)$. Therefore the minima of the
two-spinon spectrum are at the Brillouin zone edge-centers and
center, as illustrated in FIG.~\ref{fig:pispinspectrum}. The
distinction from the analogous diagram for the zero-flux state in
FIG. ~\ref{fig:zerospinspectrum} indicates that the $\pi$-flux state
is really a distinct state from zero-flux state. 

The critical quantum parameter for $\pi$-flux state is found to be
$\kappa_c\approx 0.75$. This large $\kappa_c$ makes the $\pi$-flux
state a promising candidate of spin liquid especially given that in
the zero-flux case the mean field theory seemed to overestimate
magnetic order. Even in the event that magnetic order is present, it
is expected to be weak, and the presence of a proximate spin liquid
should be apparent at finite temperatures and energies as discussed
in \cite{IsakovSenthilKim}. However, for the nearest-neighbor
antiferromagnet, the zero-flux state always has lower mean field
energy, which is consistent with the general argument of
flux-expulsion by Tchernyshyov, {\it et. al.}
\cite{Tchernyshyov:GreedyBoson}. Later we will see that $\pi$-flux
state can be stabilized as the global minimum of the mean field
theory, if other terms such as next neighbor couplings or ring
exchange are present in the Hamiltonian.

\subsection{Spin Configurations from Spinon Condensates}
When quantum parameter $\kappa$ goes beyond its critical value the
spinon dispersion becomes zero at several $\veck$-points. Bosons
condense at those $\veck$-points, namely, the spinon field
$\spinon(\veck)$ receives expectation value in ground state. We
briefly analyze the structure of condensates and the corresponding
spin configurations for the zero-flux state, where we follow
reference \cite{Sachdev:KagomeAndTrigLatt}, and then apply the same
analysis to the $\pi$-flux state on the triangular lattice.

{\em The zero-flux state}, has spinon minima at
$\veck_c=(2\pi/3,2\pi/3)$. At this point,  $D(\veck_c)$ of equation
(\ref{Dzeroflux}) has an eigenvector with zero eigenvalue 
\be
\spinon_1=\begin{pmatrix}\im,&1\end{pmatrix}^T\nn 
\ee ($^T$ is transposition). When $\kappa$
just exceeds $\kappa_c$, a condensate at $\veck_c$ is obtained
$\bra\spinon(\veck_c)\ket= (\bra b_{\veck_c\uparrow} \ket\, \bra
b_{-\veck_c\downarrow}\yd \ket)=c_1\spinon_1$. Similarly,
$D(-\veck_c)$ has one eigenvector associated with zero eigenvalue
\be \spinon_2=\begin{pmatrix}-\im,&1\end{pmatrix}^T=\spinon_1^*\nn \ee
Condensate at $-\veck_c$ is
$\bra\spinon(-\veck_c)\ket=c_2\spinon_2$. Here $c_{1,2}$ are two
complex numbers.

The condensate on lattice site $\siter$ is then given by \be x\equiv
\begin{pmatrix}
\bra b_{\siter\uparrow}\ket\\
\bra b_{\siter\downarrow}\ket
\end{pmatrix}
=
\begin{pmatrix}
\im c_1,&-\im c_2\\
c_2^*,&c_1^*
\end{pmatrix}
\begin{pmatrix}
e^{\im \veck_c\cdot\siter}\\
e^{-\im \veck_c\cdot\siter}
\end{pmatrix}\nn
\ee and the ordered magnetic moment is
$\mathbf{S}(\siter)=(1/2)x\yd\mathbf{\pauli}x$.

This can be readily shown to give rise to the 120 degree classical
Neel state, in which the classical spin is \be
\mathbf{S}(\siter)=\mathbf{n}_1 \cos(\mathbf{Q}\cdot
\siter)+\mathbf{n}_2 \sin(\mathbf{Q}\cdot \siter) \ee where
$\mathbf{n}_{1,2}$ are orthogonal to each other and have the same
length as the classical spin \be
\mathbf{n}_1^2=\mathbf{n}_2^2=\mathbf{S}^2 \ee
$\mathbf{Q}=\pm(2\pi/3,2\pi/3)$ is the wavevector of Brillouin zone
corner, represented as small triangles in FIG.~\ref{fig:BZ}.

{\em For $\pi$-flux state}, the critical spinon wavevector occurs at
$\veck_c=(\pi/2,-\pi/2)$. At this point,, $D(\veck_c)$ of equation
(\ref{equ:DPmatrix}) has two eigenvectors associated with zero
eigenvalue
\begin{eqnarray*}
\spinon_1&=&\begin{pmatrix}\frac{1-\im}{\sqrt{3}},&-\frac{\im}{\sqrt{3}},&0,&1\end{pmatrix}^T\\
\spinon_2&=&\begin{pmatrix}\frac{\im}{\sqrt{3}},&-\frac{1+\im}{\sqrt{3}},&1,&0\end{pmatrix}^T
\end{eqnarray*}
The condensate at $\veck_c$ can in general be parameterized as
$\bra\spinon(\veck_c)\ket=c_1\spinon_1+c_2\spinon_2$. Similarly,
$D(-\veck_c)$ has two eigenvectors associated with zero eigenvalue
\begin{eqnarray*}
\spinon_3&=&\begin{pmatrix}\frac{1+\im}{\sqrt{3}},&\frac{\im}{\sqrt{3}},&0,&1\end{pmatrix}^T=\spinon_1^*\\
\spinon_4&=&\begin{pmatrix}-\frac{\im}{\sqrt{3}},&-\frac{1-\im}{\sqrt{3}},&1,&0\end{pmatrix}^T=\spinon_2^*
\end{eqnarray*}
and the condensate at $-\veck_c$ is parameterized as:
$\bra\spinon(-\veck_c)\ket=c_3\spinon_3+c_4\spinon_4$. Here
$c_{1,2,3,4}$ are four complex numbers.

Returning to real space, the spinon condensate on sublattice
$\sublatone$ is \be
\begin{split}
x_{\sublatone}\equiv&
\begin{pmatrix}
\bra b_{\sublatone\siter\uparrow}\ket\\
\bra b_{\sublatone\siter\downarrow}\ket
\end{pmatrix}\\
=&
\begin{pmatrix}
\frac{1-\im}{\sqrt{3}}c_1+\frac{\im}{\sqrt{3}}c_2,&\frac{1+\im}{\sqrt{3}}c_3-\frac{\im}{\sqrt{3}}c_4\\
c_4^*,&c_2^*
\end{pmatrix}
\begin{pmatrix}
e^{\im \veck_c\cdot\siter}\\
e^{-\im \veck_c\cdot\siter}
\end{pmatrix}
\end{split}
\nn \ee while the spinon condensate on sublattice $\sublattwo$ is
\be
\begin{split}
x_{\sublattwo}\equiv&
\begin{pmatrix}
\bra b_{\sublattwo\siter\uparrow}\ket\\
\bra b_{\sublattwo\siter\downarrow}\ket
\end{pmatrix}\\
=&
\begin{pmatrix}
-\frac{\im}{\sqrt{3}}c_1-\frac{1+\im}{\sqrt{3}}c_2,&\frac{\im}{\sqrt{3}}c_3-\frac{1-\im}{\sqrt{3}}c_4\\
c_3^*,&c_1^*
\end{pmatrix}
\begin{pmatrix}
e^{\im \veck_c\cdot\siter}\\
e^{-\im \veck_c\cdot\siter}
\end{pmatrix}
\end{split}
\nn
\ee

The ordered magnetic moment on sublattice $p$ is $(1/2)\left
(x_{p}\right )\yd\mathbf{\pauli}x_{p}$ for $p=\sublatone,\sublattwo$
respectively and is present at wavevectors at the mid points of the
Brillouin Zone edges shown via black hexagons in FIG.~\ref{fig:BZ}.

Under the constraint that the magnitude of condensate is uniform
(which is required to have uniform magnitude for the ordered moments
), we find a continuum of four-sublattices ordered states, which is
consistent with previous classical energy analysis on triangular
lattice magnets with interactions that favor order at these
wavevectors \cite{Korshunov:ChiralPhaseTrigAFMwithRing}. The
classical spin in the four-sublattices states is \be
\mathbf{S}(\siter)=\mathbf{n}_1 (-1)^{r_1}+\mathbf{n}_2
(-1)^{r_2}+\mathbf{n}_3 (-1)^{r_1+r_2} \ee where
$\mathbf{n}_{1,2,3}$ are three vectors orthogonal to each other, and
\be \mathbf{n}_1^2+\mathbf{n}_2^2+\mathbf{n}_3^2=\mathbf{S}^2. \ee

Generally speaking this configuration is non-coplanar if
$|\mathbf{n}_{1,2,3}|$ are all nonzero. However, previous analysis
confirmed the existence of \quo order from disorder\quo phenomena, namely,
quantum or thermal fluctuation will lift the accidental degeneracy
and favor a collinear state in large S models with next nearest
neighbor antiferromagnetic exchange \cite{Jolicoeur:ClassicalEnergy,
Chubukov:OrderFromDisorderAFMTrigJ1J2}, which corresponds to the
case that only one of $\mathbf{n}_{1,2,3}$ is nonzero.

It has been argued that ring exchange will favor
a non-coplanar state\cite{Korshunov:ChiralPhaseTrigAFMwithRing}.
The classical ground state will then be
a tetrahedral configuration with equal angle between any two of the four spins.
In the above expression, this state is realized when $\mathbf{n}_{1,2,3}$ have same magnitude.

However, on studying the ordered state with weak spinon
condensation,  the constraint that the magnitude of the spinon
condensate is uniform enforces the additional condition on the
$\mathbf{n}_{1,2,3}$, that one of the following three equations must
be satisfied
\begin{subequations}
\begin{eqnarray}
|\mathbf{n}_{1}|&=&|\mathbf{n}_{2}|+|\mathbf{n}_{3}|\\ {\rm
or}\,\,\, |\mathbf{n}_{2}|&=&|\mathbf{n}_{3}|+|\mathbf{n}_{1}|
\\{\rm or}\,\,\, |\mathbf{n}_{3}|&=&|\mathbf{n}_{1}|+|\mathbf{n}_{2}|
\end{eqnarray}
\end{subequations}

Thus the classical degeneracy is not entirely lifted at this level.
Also, neither the collinear state nor the tetrahedral state can be
obtained from weak spinon condensates from this spin liquid state -
although the wavevectors of all these ordered states are identical. 
This implies that if there is a continuous phase transition out of
the $\pi$-flux spin liquid phase, the proximate spin ordered state
is not the collinear or tetrahedral state, but these are realized
via further phase transitions on increasing the size of the spin. We
are still seeking a simple explanation for these results.

\section{Hamiltonians Stabilizing the $\pi$-flux State}\label{sec:mftheory}
The $\pi$-flux spin liquid state differs in several respects from
the zero-flux state; also, its stability against magnetic ordering
up to a relatively large quantum parameter  $\kappa_c=0.75$ (in a
nearest neighbor model) suggests its physics may be important for
understanding physical spin-1/2 models. Hence it is of interest to
ask what interactions might stabilize such a $\pi$-flux state.

A clue is provided by the fact that the lowest energy triplet
excitations of such a state are located at the midpoints of the
Brillouin zone edges. This is precisely the ordering wave-vector for
{\em classical} spins on the triangular lattice in a model with
next-nearest-neighbour ($J_{2}$) antiferromagnetic interactions in
addition to nearest-neighbour antiferromagnetic interactions
($J_{1}$) in the range $0.125J_{1}<J_2<J_{1}$\cite{Jolicoeur:ClassicalEnergy}. It is conceivable
that on reducing the size of the spin in this model from infinity
(the classical limit) to S=1/2, the system enters a spin liquid
state described by the $\pi$-flux state, although strictly the
quantum parameter for this case is $\kappa=1$.

A second possibility arises from ring exchange - it is well known
that there are several realization of S=1/2 quantum antiferromagnets
on the triangular lattice where ring exchange plays an important
role. These include monolayers of $^3$He adsorbed on graphite
\cite{He3}, and the organic quantum antiferromagnet
$\kappa$-(ET)$_2$ Cu$_2$(CN)$_3$ \cite{Kanoda}. Also, exact
diagonalization studies of the triangular lattice S=1/2 magnet with
ring exchange has uncovered unusual properties
\cite{Misguich:ExactDiagTrigLattMultipleSpinPhaseDiagram}. In a
recent variational study with fermionic mean field states by
Motrunich \cite{Motrunich:ProjectedFermiSea} it was found that ring
exchange of the sign that arises naturally with electrons, favors a
state with zero flux. Here we will investigate the effect of ring
exchange on the Schwinger boson states - in particular for which
sign of the exchange one might favor the $\pi$-flux state. We find
that including four spin ring exchange with the sign as obtained
from the Hubbard model stabilizes the $\pi$-flux state.

Here we give some physical arguments as to why next-neighbor
antiferromagnetic interactions and the ring term will stabilize
$\pi$-flux state over the zero-flux state.

\subsection{Effect of Next Neighbour Interactions $J_2$:}

For zero-flux state, nearest-neighbor amplitudes $B_{1}$ and
$A_{1}$, and next-nearest-neighbor amplitudes $B_{2}$ are nonzero,
while the next-nearest-neighbor amplitude $A_{2}$ is zero. Note that
$B_{2}$ is nonzero even if next-nearest-neighbor coupling $J_{2}$ is
zero. \be
\begin{split}
\frac{E\zeroindex}{\mathrm{bond}\cdot J_{1}}=&\phantom{+}\left (B\zeroindex_{1}\right )^2-\left (A\zeroindex_{1}\right )^2\\
&+\frac{J_{2}}{J_{1}} \left (B\zeroindex_{2}\right )^2
\end{split}
\nn
\ee

In contrast, for the $\pi$-flux state, $A_{1}$ and $A_{2}$ are
nonzero, while $B_{1}$ and $B_{2}$ are zero.  Note, $A_{2}$ is
nonzero even if $J_{2}$ is zero. \be
\frac{E\piindex}{\mathrm{bond}\cdot J_{1}}=-\left
(A_{1}\piindex\right )^2-\frac{J_{2}}{J_{1}} \left
(A_{2}\piindex\right )^2\nn \ee

It is readily shown with the self consistent values of the mean
field amplitudes that for $J_{2}=0$, $E\zeroindex<E\piindex$, i.e.
the zero-flux state is preferred at the mean field level. However,
if we increase $J_{2}$ from zero, we do not expect the ansatz to
change rapidly, thus the energy for zero-flux state will have a
positive slope with respect to $J_{2}$, while the $\pi$-flux state
has a negative slope. We may then expect a first-order transition
between the two states. This is consistent with previous analysis of
$J_{1}-J_{2}$ model in the opposite limit of semiclassical spins,
where a transition between the 120 degree state (the magnetically
ordered analog of the zero-flux state and the colinear state, the
magnetically ordered analog of the $\pi$-flux state is found on
increasing $J_2$ \cite{Chubukov:OrderFromDisorderAFMTrigJ1J2,
Deutscher:SpinWaveJ1J2}. From the classical energy
analysis\cite{Jolicoeur:ClassicalEnergy}, we know that in large-$S$
limit $\alpha=1/8$ is the critical point between different ordered
states with ordering wavevectors at the BZ corners vs. BZ edge
midpoints. Thus one might expect that a moderate
next-nearest-neighbor term ($1/8<\alpha<1$) will stabilize the
$\pi$-flux state over the zero-flux state in the quantum limit,
since these states have short range order (two spinon minima) at
precisely these wavevectors. We now provide estimates for the phase
boundary (the critical ratio $\alpha= J_2/J_1$) between the
zero-flux and $\pi$-flux spin liquid phases as a function of the
quantum parameter $\kappa$.

{\em Small $\kappa$ analysis:} The competition between different
spin liquid states can be studied analytically in the limit of
small-$\kappa$\cite{Tchernyshyov:GreedyBoson}. In this case the magnitude of the ansatz $A_{ij}$
and $B_{ij}$ are small. Then, we can develop a series expansion of
the self-consistent equations in $\kappa$, and find solutions.

First, we use this expansion to look at the $J_1$-$J_2$ model, with
$\alpha\equiv J_2/J_1<1$ and sufficiently small quantum parameter
$\kappa$.



The energy difference between the zero-flux and the $\pi$-flux state
in this limit is: \begin{eqnarray}\nonumber \frac{
E\zeroindex-E\piindex }{\mathrm{bond}\cdot J_{1}}&=&
-(1-\alpha)\left [ \frac{\kappa}{6} \right
]^2+\frac{2\alpha}{1-\alpha}\left [ \frac{\kappa}{6} \right ]^3 \\
\nonumber &&+\left (\frac{87}{4}+23\alpha+ 72\alpha^2\right )\left [
\frac{\kappa}{6} \right ]^3\\ &&+O(\kappa^4)
\end{eqnarray}

From the above expressions for mean field energy difference, we can
find the transition point between the zero-flux and $\pi$-flux
states to lowest order in $\kappa$. 
\be 
\alpha_c=1-\sqrt{\kappa/3}\label{equ:criticalalpha}
\ee In the limit of vanishing $\kappa$ the zero-flux state is
preferred until the next-neighbor bond is stronger than the nearest
neighbor bond. However, this critical ratio decreases rapidly with
increasing $\kappa$.

In the small $\kappa$ limit one can ignore the spin carrying
excitations. Then, going beyond the mean field theory the relevant
degrees of freedom are the gauge excitations, which in this case
correspond to those of an Ising gauge theory. The distinction
between zero and $\pi$-flux states in this limit we believe has to
do with the the sign of plaquette energy term that penalizes Ising
vortex excitations in the case of the zero-flux state, but prefers a
ground state with a uniform background \quo flux\quo of such vortices in
the case of the $\pi$-flux state.

\begin{figure}[h]
\includegraphics{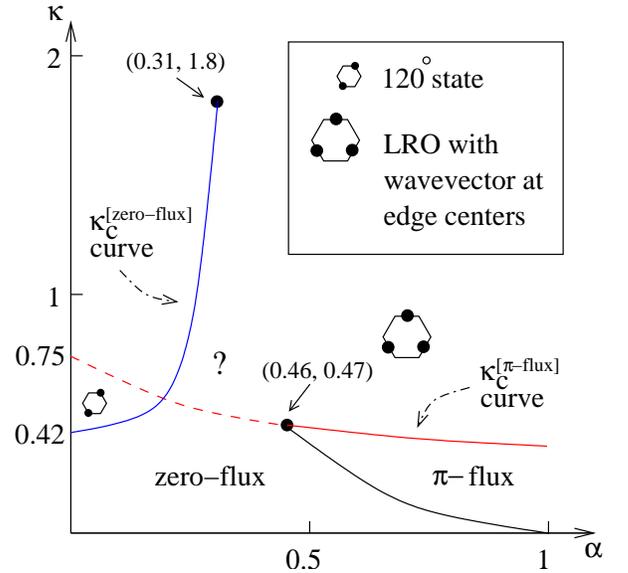}
\caption{\label{fig:meanfield_phase} (color online) The mean field
phase diagram of the triangular lattice antiferromagnet with nearest
neighbor  $J_1$ and next-neighbor $J_2$ interactions, as a function
of the ratio $\alpha=J_2/J_1$ and quantum parameter $\kappa$. The
diagram is a quantitatively reliable solution of mean field
equations in the region near to and below the red line (dashed and
solid) where spinon condensation of the $\pi$-flux state occurs.
Just above the solid red line, magnetic ordering at the Brillouin
zone M-point occurs, while below it the $\pi$-flux spin liquid phase
is obtained. The black line is the first-order phase boundary
between the zero-flux and $\pi$-flux spin liquid phases, while the
blue line represents the onset of 120 degree magnetic order. The
quick rise of this line to large values of $\kappa$ is due to
frustration of the 120 degree Neel order by next neighbor
interactions. If uninterrupted by magnetic order (of the M type)
this may open a window of spin liquid for physical spin values in
this model.}
\end{figure}

{\em Numerical Solution:} The mean field equations can be
numerically solved to obtain the competition between the zero-flux
and $\pi$-flux states and states with magnetic order. This is shown
in FIG.~\ref{fig:meanfield_phase}. The critical value
$\kappa_c\piindex$ at which spinon condensation occurs for the
$\pi$-flux state is shown by the dashed and solid red line. If the
$\pi$-flux state is stabilized, then above this line spinon
condensation leads to magnetic order at the B.Z. edge centers
(M-points). There are a number of distinct ordering patters
consistent with this ordering wavevector - which makes an analysis
of the ordered phase a delicate one; and we do not attempt it here.
Therefore, the phase diagram shown in FIG.~\ref{fig:meanfield_phase} 
is strictly speaking only to be trusted
below this $\kappa_c\piindex$ line(red line). However, certain extrapolations
above this line can be made with some degree of confidence.

The first-order phase boundary  between the $\pi$-flux and zero-flux
spin liquid states is shown by the black line in FIG.~\ref{fig:meanfield_phase}. 
At small $\kappa$ it approaches
$\alpha=1$ via the analytic form derived previously \ref{equ:criticalalpha}. It intersects
the spinon condensation line for the $\pi$-flux state at
$\alpha=0.46,\ \kappa=0.47$.

The zero-flux state is stabilized at smaller values of $\alpha$, and
sufficiently small quantum parameter $\kappa<\kappa_c\zeroindex$, where the
critical quantum parameter for spinon condensation in this state is
shown by the blue line. Spinon condensation is expected to lead to
the 120 degree state on crossing this line\cite{incommensurate}. Again, this is most reliably established if the
ordered states with wavevectors at the M-points (M-states) are not
in the picture, which is the case below the dashed red line in
FIG.~\ref{fig:meanfield_phase}. An interesting aspect of the phase
diagram is the large values of $\kappa_c\zeroindex \approx 1.8$ that arise, if
the M-states are neglected. This is believed to be a result of
frustration of the 120 degree Neel order by next neighbor
interactions. While some part of this spin liquid region in the
phase diagram above $\kappa_c\piindex$ might be occupied eventually by
magnetically ordered M-states, the existence of zero-flux spin
liquid states for fairly large values of the quantum parameter is
likely to persist, making the $J_1$-$J_2$ model an interesting
candidate for spin liquid physics that deserves further attention.

\subsection{Effect of Ring Exchange}
In addition to the exchange interaction between a pair of spins,
multi-spin interactions are also present in insulators with
non-negligible charge fluctuations. While for spin one-half systems,
the three spin interaction can be expressed in terms of two spin
exchange, a new term is generated when considering four spin
interactions. The four spin ring exchange term is:
\begin{eqnarray}
H_{Ring} &=& J_{Ring}\sum{(P_4+h.c.)}\\
P_4 &=& b_{i\delta}\yd b_{i\alpha}\nd b_{j\alpha}\yd b_{j\beta}\nd
b_{k\beta}\yd b_{k\gamma}\nd b_{l\gamma}\yd b_{l\delta}\nd
\end{eqnarray} where $P_4$ is
the four spin permutation operator, which is written out
conveniently in terms of the Schwinger boson operators as above. The
sites $i,j,k,l$ form a rhombus and $\alpha,\,\beta,\,\gamma,\,
\delta \in \{\uparrow,\, \downarrow\}$ are spin indices. The above
bosonic spinon representation of the ring-exchange may be checked by
expanding it in terms of the spin operators in the spin-1/2 case and
verifying it has the required form \cite{Spin}. A more efficient way
to prove this identity will be to establish a correspondence with
the fermionic spinon representation of ring exchange which is done
later in this section.


From the $t/U$ expansion of the Hubbard model
\cite{MacDonald:tUexpansion}, it is found that $J_{Ring}=20 t^4/U^3$
and is positive. In the following, when we consider ring exchange
term we neglect the next-nearest-neighbor term, and vise versa.
Evaluating the  ring exchange term in the mean field ground state,
one find that the result contains:
\begin{equation}
\langle H_{Ring}\rangle_{MF} = 8J_{Ring}(A_{ij}\nd A_{jk}^*A_{kl}\nd
A_{li}^*+c.c) + \ldots
\end{equation}
which is the term we used to define flux. The other terms are
small, especially in the small-$\kappa$ regime. This term clearly
has different signs for zero-flux and $\pi$-flux states, and favors
the $\pi$-flux states when $J_{Ring}$ is positive. Thus we also
expect a possible transition from zero-flux state to $\pi$-flux
state when we tune ring exchange coupling.

It is well known that the ring exchange term, derived from the Hubbard model, has positive
(negative) coupling constant when expressed in terms of a
permutation operator, if the loop length is even (odd)\cite{Thouless:}. 
This can be seen as follows.
Since the ring exchange term arises from virtual electron hopping
process lowering the energy, the effective term generated is
$-K_{L}[c_{1\sigma_{L}}\yd c_{L\sigma_{L}}\nd c_{L\sigma_{L-1}}\yd
c_{L-1\sigma_{L-1}}\nd \cdots c_{2\sigma_{1}}\yd
c_{1\sigma_{1}}\nd+\mathrm{h.c.}]$ where $c_{i\sigma}$ are electron
annihilation operators with site index $i$ and spin index $\sigma$,
and $K_{L}>0$. Clearly this term includes a permutation operator for
length $L$ loop $(-1)^{L-1}(-K_{L})c_{1\sigma_{L}}\yd
c_{1\sigma_{1}}\nd c_{2\sigma_{1}}\yd c_{2\sigma{2}}\nd \cdots
c_{L\sigma_{L-1}}\yd c_{L\sigma_{L}}\nd +\mathrm{h.c.}$. So for $L$
even we have $+K_{L}(P_{L}+\mathrm{h.c.})$.

While the above discussion has been phrased in terms of fermionic
spinons, we can now translate those observations into the bosonic
spinon language. It turns out that the permutation operator has the
same form and sign in terms of both fermion (electron) operators and
Schwinger boson operators. We give two arguments in support of this
below. First, notice that spin $1/2$ operators has the same
representation in Schwinger boson and \quo Schwinger fermion\quo 
(electrons) scheme. Assume that spin indices $\alpha,\ \beta$ take
values of +1(spin up) and -1(spin down). For $S=1/2$ system we have
$c_{i\alpha}\yd c_{i\beta}\nd=(2S+2\alpha
S^{z})\delta_{\alpha,\beta}+\left
[(1+\alpha)S^{+}+(1-\alpha)S^{-}\right
]\delta_{\alpha,-\beta}=b_{i\alpha}\yd b_{i\beta}$. Therefore we can
simply replace fermion operator $c$ by Schwinger boson operator $b$
in the permutation operator. The second argument is to use
Jordan-Wigner transformation to change electronic operators into
boson. After replacing fermionic operators with bosons the
Jordan-Wigner string operators becomes a constant under the
constraint of constant($2S$) particles on each site.

Now we evaluate the above ring exchange operator for longer length
loops in the Schwinger boson mean field states. We focus on even
length loops since we expect the dominant amplitudes to be given by
the $A_{ij}$s, which can only give rise to even length loops if they
are the only amplitudes that count. The dominant term is $2\cdot
2^{L/2}\cdot(A_{1,2}\nd A_{2,3}^* \cdots A_{L-1,L}\nd
A_{L,1}^*+\mathrm{h.c.})$. This is not exactly the term defining
flux in the loop\cite{Tchernyshyov:GreedyBoson}, which has an
additional minus sign for each $A^*$. Namely the result is $2\cdot
2^{L/2}\cdot (-1)^{L/2} (\mathrm{term\ defining\ flux+h.c.})$. It is
then clear that for $L=4,8,12,\dots$ the ring exchange term favors
$\pi$-flux in the loop, while for $L=6,10,14,\dots$ it favors zero
flux. It is amusing to contrast this with the result for mean field
theories based on a fermionic spin representation, where ring
exchange favors states with zero-flux
\cite{Motrunich:ProjectedFermiSea}. The difference has to do with
the different definitions for flux that naturally appear in these
two theories.


{\em Small $\kappa$ Analysis:} As argued previously, raising the
value of $J_{Ring}/J_1$ above a critical value, favors the $\pi$
flux state over the zero-flux state. We now study the effect of ring
exchange term by evaluating it in the zero-flux and $\pi$-flux
states, in the limit of small $\kappa$ where an analytic solution
may be obtained. Another benefit for going to small-$\kappa$ limit is 
that larger-ring exchange effect ({\it e.g.} length-six ring) is controlled by 
small parameter $\kappa$ as $\kappa^{\mathrm{length/2}}$ according to Tchernyshov, {\it et. al.}
\cite{Tchernyshyov:GreedyBoson}, therefore only the energetics of 
length-four ring, namely the rhombus, is important.

The procedure described above is of course only the lowest order term, but
we expect it to give us the correct qualitative behavior. For
zero-flux state, the contribution from the ring exchange term is
$(\kappa^2/6) J_{Ring}$ per rhombus to lowest order. For $\pi$-flux
state, it is $-(\kappa^2/18) J_{Ring}$ per rhombus to lowest order.
Therefore, a positive $J_{Ring}$ clearly favors the $\pi$-flux
state. Since for every site, there are three bonds and three rhombi
on average, we can find the transition point in the
$\kappa\rightarrow 0$ limit to be : \be [J_{Ring}/J_{1}]_c=1/8.\ee
Several other theories predict a transition for the triangular
lattice antiferromagnet while tuning this ratio $J_{Ring}/J_{1}$,
although the phases appearing may be different. Motrunich's
fermionic variational method \cite{Motrunich:ProjectedFermiSea}
predicts a transition from a \quo $\pi$-flux\quo fermionic mean field state
to \quo zero-flux\quo mean field state when $J_{Ring}/J_{1}$ increases over
about $0.35$. A previous bosonic mean field phase diagram
\cite{Misguich:SBMultipleSpinPhaseDiagram} predict a transition from
120 degree Neel-ordered state to collinear-ordered state when this
ratio increases over $1/3$. It is interesting that we found a
similar transition between two disordered spin liquid state for
sufficiently small $\kappa$, with short ranged order at precisely
these two wavevectors, as this ratio is increased.

{\em Numerical Solution:} For finite $\kappa$, the numerical
solution of the mean field equations have been obtained for the
nearest-neighbor model below the critical value $\kappa_c$ for both
states. We can make a lowest order perturbative treatment of
next-nearest-neighbor and ring exchange terms, namely we evaluate
these two terms in the mean field ground state of nearest-neighbor
model and thus obtain the transition point. We found approximate
critical value of $\alpha\equiv J_{2}/J_{1}$ and $J_{Ring}/J_{1}$,
listed in TABLE~\ref{tab:fintekapparesult}.
\begin{table}[h]
\begin{center}
\begin{tabular}{|l||l|}
\hline
$\kappa$ &$J_{Ring}/J_{1}$\\
\hline
0.05&0.126\\
0.10&0.127\\
0.20&0.128\\
0.30&0.129\\
\hline
\end{tabular}
\caption{Approximate finite $\kappa$ result for critical
$J_{Ring}/J_{1}$ at which the zero-flux state gives way to the
$\pi$-flux state on the triangular
lattice.}\label{tab:fintekapparesult}
\end{center}
\end{table}


\section{Spin Liquid States on the Kagom\'e Lattice}
\label{sec:KagomePSG}
\begin{figure}[h]
\includegraphics{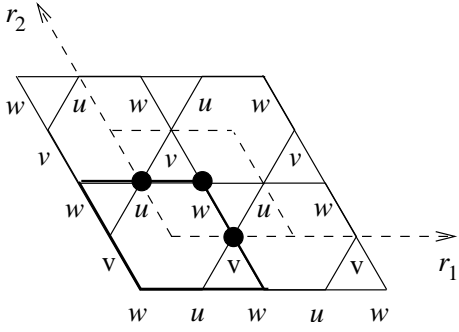}
\caption{\label{fig:Kagome} Coordinates system of Kagome lattice.
Rhombus enclosed by dash lines is the unit cell at origin. Three
sites with black dots form the basis of lattice.
Black rhombus is an example of length-eight loops.}
\end{figure}

The Kagom\'e lattice has three sites in one unit cell, labeled by
$\sublatone,\ \sublattwo,\ \sublatthree$ (see
FIG.~\ref{fig:Kagome}).

We label a unit cell by $\siter=(r_1,r_2)$ and label a site by
$(r_1,r_2)_{p}$ where $p=\sublatone,\sublattwo,\sublatthree$.

\subsection{Projective Symmetry Group Analysis of the Kagom\'e Lattice}
We would like to obtain the possible symmetric spin liquid states on
the Kagom\'e lattice described within the Schwinger boson approach.
Again, a Projective Symmetry Group analysis will be carried out just
as was done on the triangular lattice. A gauge transformation
$\PSG{}$ is then described by three phase functions
$\Gphase{\sublatone}(\siter), \Gphase{\sublattwo}(\siter),
\Gphase{\sublatthree}(\siter)$.

For simplicity we will consider ansatz with $A_{ij}=\frac12
\epsilon_{\sigma\sigma'}\langle b_{i\sigma}b_{j\sigma'}\rangle$
only. These amplitudes are naturally present for predominantly
anti-ferromagnetic interactions, and can be strictly justified in
large-N limit \cite{Sachdev:KagomeAndTrigLatt}. Then, due to the
frustrated nature of the Kagom\'e lattice, the Invariant Gauge Group
is $Z_2$, generated by \be
\Gphase{\sublatone}(\siter)=\Gphase{\sublattwo}(\siter)=\Gphase{\sublatthree}(\siter)=\pi
\nn \ee thus we will study $Z_2$ spin liquid states within the
Schwinger boson approach.

In the above coordinates system the space group is generated by two
translations $\Tone,\ \Ttwo$, reflection $\mirror$ and 60 degree
rotation $\rotsix$. The first two generators preserve sublattices.
The reflection exchanges $\sublatone$ and $\sublattwo$ sublattices.
The rotation is a cyclic permutation of three sublattices
$\sublatone\rightarrow\sublattwo,\
\sublattwo\rightarrow\sublatthree,\
\sublatthree\rightarrow\sublatone$.

For each space group generator $X$ we can associate a gauge
transformation described by three phase functions
$\Gphase{X,\sublatone}(\siter), \Gphase{X,\sublattwo}(\siter),
\Gphase{X,\sublatthree}(\siter)$.

The procedure of solving the for the different PSGs allowed by the
algebraic constraints imposed by relations between symmetry elements
is parallel to triangular case, except that one has to keep track of
the index of sublattices. It turns out that the three phase
functions can be chosen to be identical. The solution of algebraic
PSG (details in Appendix~\ref{asec:PSGKagome}) is then {\em
identical} to the triangular lattice case except that the phase
functions have one more index.
\begin{subequations}
\begin{eqnarray}
\Gphase{\Tone,p}(r_1,\ r_2)&=&0\label{equ:PSGKagomephiT1}\\
\Gphase{\Ttwo,p}(r_1,\ r_2)&=&p_1\pi r_1\label{equ:PSGKagomephiT2}\\
\Gphase{\mirror,p}(r_1,\ r_2)&=&p_2\pi/2+p_1\pi r_1 r_2 \label{equ:PSGKagomephimirror}\\
2 \Gphase{\rotsix,p}(r_1,\ r_2)&=&p_3\pi+p_1\pi
r_2(r_2\!-\!1\!+\!2r_1)\label{equ:PSGKagomephirotsix}
\end{eqnarray}
\end{subequations}
where $p=\sublatone,\sublattwo,\sublatthree$. Thus, in the case of
the Kagom\'e lattice too, there are not more than 8 symmetric $Z_2$
spin liquid states, corresponding to the two values that each of the
$p_1,\,p_2,\,p_3$ can take on.

If we now specialize to states that have  non-vanishing
nearest-neighbor amplitudes $A_{ij}$, we have more realizations of
the algebraic PSG. The only additional restriction({\it c.f.}
SEC.~\ref{ssec:realization}) is that bond
$(0,0)_{\sublatone}\rightarrow (0,0)_{\sublatthree}$ can be related to bond
$(0,0)_{\sublattwo}\rightarrow (0,0)_{\sublatthree}$ by both reflection and 60
degree rotation. This gives one constraint $p_2=1-p_3$. In all these
ansatz the amplitudes $A_{ij}$ are real and of uniform magnitude.

\subsection{Spin Liquid States on the Kagom\'e}
The four realizations are listed below. We will use the same
convention of $\veck$-space coordinates system as triangular case by
taking the lattice constant as unity (the length of nearest-neighbor
bond is 1/2). It is convenient to divide the four states into two
pairs, two with $p_1=0$ and two with $p_1=1$. The first condition
implies that the generators of translation commute, the analog of
the zero-flux state, while the second condition implies a flux of
$\pi$ in a unit cell. It will turn out that the two zero-flux states
correspond to the two states discussed earlier by Sachdev in the
context of the Kagom\'e lattice, while the two $pi$-flux states have
not previously been discussed.

\begin{figure}[h]
\includegraphics{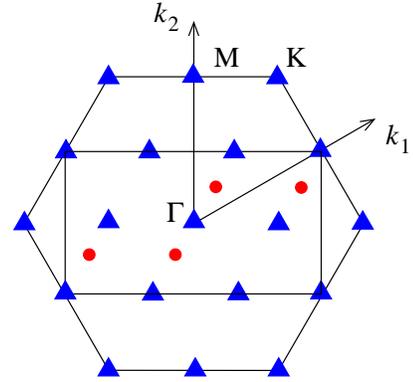}
\caption{\label{fig:BZKagome}(Color online). $\veck$-space of
Kagom\'e lattice. The large hexagon is Brillouin zone of original
Kagom\'e lattice and the two $p_1=0$ ansatz. The large rectangle is
Brillouin zone of the two $p_1=1$ ansatz, which enclose a flux of
$\pi$ in their unit cells. Red dots are the minima of the spinon
dispersion for both $p_1=1$ states within the reduced Brillouin
zone. Blue triangles are the locations of the minima of the lower
edge of the two-spinon spectrum for both $p_1=1$ states. }
\end{figure}

{\em The Zero-Flux States:}

1.  The first state we consider is characterised by $p_1=0,\ p_2=0,
p_3=1$. This is the large-N ground state identified by Sachdev for
the nearest neighbor antiferromagnet, which he called $Q_1=-Q_2$
state\cite{Sachdev:KagomeAndTrigLatt}.

This state has zero-flux in both length-six hexagon and length-eight
rhombus, hence we will also refer to it as [$0$Hex,$0$Rhom]. Its
critical quantum parameter $\kappa_c\approx 0.54$. The spinon
dispersion has minima at corners of Brillouin zone
$\veck=\pm(2\pi/3,2\pi/3)$, and so does the two-spinon spectrum.
After spinon condensation it gives rise to the
$\sqrt{3}\times\sqrt{3}$ magnetically ordered classical ground
state.

2. The second state we consider is characterized by $p_1=0,\ p_2=1,
p_3=0$. This is another state considered by Sachdev and was called
the $Q_1=Q_2$ state\cite{Sachdev:KagomeAndTrigLatt}.

This state has $\pi$ flux in length-six hexagon and zero-flux in
length-eight rhombus, hence we will also refer to it as
[$\pi$Hex,$0$Rhom]. Its critical quantum parameter $\kappa_c\approx
0.50$. The spinon dispersion has minima at centers of Brillouin zone
$\veck=(0,0)$, so does the two-spinon spectrum. After spinon
condensation it gives rise to the translational invariant classical
ground state.

{\em The $\pi$-Flux States:}

3. The third state we consider is characterized by $p_1=1,\ p_2=0,
p_3=1$. The ansatz is given in FIG.~\ref{fig:Kagome101state} in the
Appendix~\ref{asec:PSGKagome} and has not been considered before.
Due to the presence of flux in the unit cell, a doubled the unit
cell needs to be considered for the spinons.

This state has $\pi$ flux in both length-six hexagon and
length-eight rhombus, hence we will also refer to it as
[$\pi$Hex,$\pi$Rhom]. Thus, in anearest neighbor model it is
expected to have the highest energy of the four states for same
$\kappa$ in the small $\kappa$ regime.

Its critical quantum parameter is fairly large, $\kappa_c\approx
0.93$. The minima of spinon dispersion is at
$\veck=\pm(\pi/6+n\pi,\pi/6+m\pi)$, where $n,\ m$ are integers(see
FIG.~\ref{fig:BZKagome}). Thus the two-spinons spectrum lower edge
has minima at $\pm(\pi/3+n\pi,\pi/3+m\pi)$ and $(n\pi,m\pi)$. In
addition to the Brillouin zone center, corners and edge centers,
minima are also present at the halfway points to the corners.

\begin{figure}[h]
\includegraphics{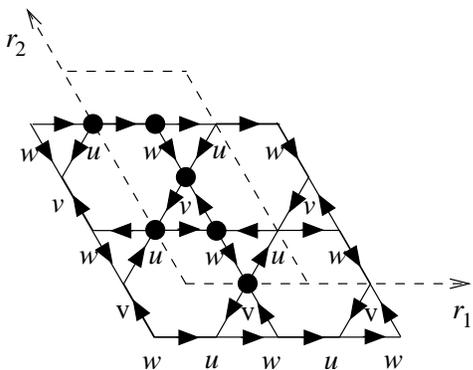}
\caption{\label{fig:Kagome110state} Ansatz of the $p_1=1,\ p_2=1,\
p_3=0$ state. Presence of $\pi$ flux in the Kagome lattice unit cell
leads to unit cell doubling, which is now defined by the dashed
rhombus. Six sites with black dots form the basis. All
nearest-neighbor $A_{ij}$ are real and of the same magnitude, and
are positive in the directions shown. This spin liquid ansatz has
zero flux in the length six hexagonal loops but $\pi$ flux in the
length eight loops such as the rhombus, hence [$0$Hex,$\pi$Rhom]. It
has a large critical quantum parameter $\kappa_c=2.0$.}
\end{figure}

4. The last state we consider is $p_1=1,\ p_2=1, p_3=0$. The ansatz
is given in FIG.~\ref{fig:Kagome110state} and has not been
considered before.

This state has zero flux in length-six hexagon and $\pi$ flux in
length-eight rhombus,hence we will also refer to it as
[$0$Hex,$\pi$Rhom] .

It is potentially the most interesting of the states considered so
far since its critical quantum parameter $\kappa_c\approx 2.0$ is
greater than unity. This means that this state is very likely to be
a symmetric spin liquid state for spin-1/2 system if it can be
realized. The large value of $\kappa_c$ is a result of the
dispersion of the relevant spinon bands being nearly flat along
certain directions in the Brillouin zone.

However, since this state has $\pi$ flux in the length-eight
rhombus, it is, by the general argument of flux-expulsion by
Tchernyshyov {\it et. al.}\cite{Tchernyshyov:GreedyBoson} expected
to have a higher mean field energy than Sachdev's ground state if we
consider a pure nearest-neighbor Heisenberg model. In the small
$\kappa$ limit the energy difference will be the order of
$\kappa^4$. However, it must also be kept in mind that mean field
energetics is not precisely the same as the true energetics of the
system.

Given the interesting nature of this state we can ask what
interactions are likely to stabilize it within mean field theory,
relative to the Sachdev ground state. From our previous analysis of
ring exchange we know that adding ring exchange interactions of the
sign obtained from the Hubbard model favors zero flux in loops of
length six and $\pi$ flux in loops of length eight. This is
precisely satisfied by loops in this ansatz as can be checked from
the FIG.~\ref{fig:Kagome110state}. Therefore the Kagom\'e
antiferromagnet with ring exchange may be a good system to realize a
spin liquid phase.

\begin{figure}
\includegraphics{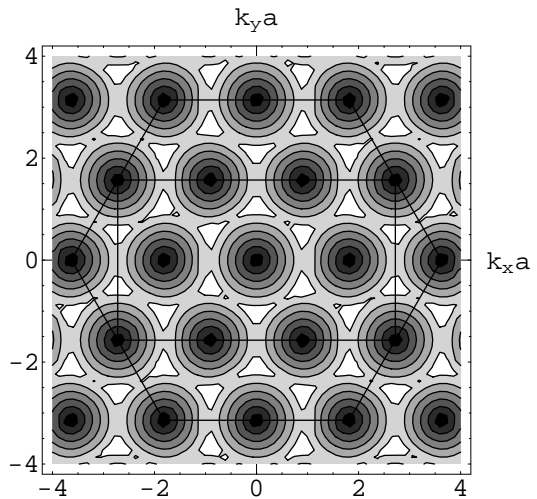}
\caption{\label{fig:twospinon101state} Two-spinons spectrum lower
edge of the $p_1=1,\ p_2=0,\ p_3=1$ state at $\kappa=\kappa_c=0.9$.
Axes are in dimensionless unit $k_{x,y}a$ where $a$ is lattice constant(two times nearest neighbor distance). 
Darker regions have lower energy.}
\end{figure}

\begin{figure}
\includegraphics{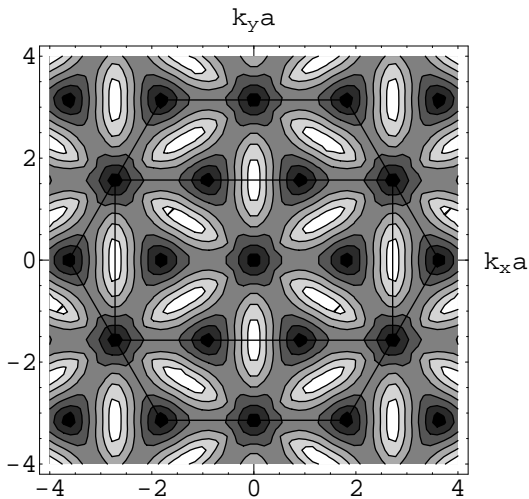}
\caption{\label{fig:twospinon110state} Two-spinons spectrum lower
edge of the $p_1=1,\ p_2=1,\ p_3=0$ state at $\kappa=\kappa_c=2.0$. 
Axes are in dimensionless unit $k_{x,y}a$ where $a$ is lattice constant(two times nearest neighbor distance). 
Darker regions have lower energy. Although the two spinon minima
occur at the same wavevector locations as in the previous $p_1=1,\
p_2=0,\ p_3=1$ state, the differences in the dispersion are
apparent.}
\end{figure}

The position of minima of spinon dispersion and two-spinons spectrum
lower edge are the same as in the previous state: $p_1=1,\ p_2=0,
p_3=1$. However, the spinon dispersions of these two states are not
identical. In particular the present $p_1=1,\ p_2=1, p_3=0$ state
has three doubly-degenerate dispersion branches while previously
discussed the $p_1=1,\ p_2=0, p_3=1$ state has six non-degenerate
dispersion branches. Other differences between these two states may
be observed in the contour plots of the lower edge of the two spinon
dispersions in FIG.~\ref{fig:twospinon110state} and FIG.~\ref{fig:twospinon101state} 
respectively. A variety of magnetically
ordered states can arise from spinon condensation in both these
cases; the magnetic states so obtained and the associated quantum
transitions would be an interesting topic for future study.

\section{Conclusions}\label{sec:conclusion}
We have extended the projective symmetry group analysis, previously
used to classify fermionic mean field states, to bosonic mean field
states of quantum antiferromagnets on the triangular and Kagom\'e
lattices. This allowed us to access fundamentally different spin liquid states and
ensure that potentially interesting states were not overlooked. On
the triangular lattice a new $Z_2$ symmetric spin liquid state
($\pi$-flux state) emerged from this analysis and was studied. 
This state is distinct from the previously studied bosonic mean field
state (zero-flux or Sachdev state), which can be seen in that they
have different spinon and two-spinons dispersions . The nature of
the ordered state resulting from spinon condensation turns out to be
more intricate in this case, as is the quantum ordering transition,
and is a topic for further study. Although this state is less
favorable from the point of view of mean field energy as compared to
the zero-flux state in the case of nearest neighbor interactions, it
is stabilized by the inclusion of moderate next nearest-neighbor or
ring exchange interactions with signs as derived from the Hubbard
model. The mean field analysis suggests that the triangular lattice
model with next neighbor antiferromagnetic interactions might be
favorable for realizing spin liquid ground states.

We have also studied $Z_2$ symmetric spin liquid states on the
Kagom\'e lattice using the PSG technique. Again we find only a few
candidate states, in particular two physically interesting
states are found which are analogous to the $\pi$-flux state on the
triangular lattice. In particular, one of these states has a
remarkably large value for the critical quantum parameter
$\kappa_c=2.0$ which implies that if realized, it would be stable
against magnetic order for physical values of the spin. We find that
ring exchange interactions would stabilize precisely this state,
making the Kagom\'e model with ring exchange an attractive model to
study in the search for spin liquid phases. Stability of these
states to other forms of order, such as valence bond solid order, is
important to understand but were not addressed in the present work.
Also, the relation between fermionic mean field states, classified
via the $SU(2)$ PSGs and the bosonic spinon mean field theories
studied in this paper with their $U(1)$ PSGs remains unclear and is
a topic for future study.

 \section{Acknowledgements}
We would like to thank O. Motrunich and C. Nayak for discussions.
A.V. was supported by A. P. Sloan Foundation Fellowship and by DOE
LDRD DEA 3664LV.
\appendix
\section{Solution to algebraic PSGs of Triangular Lattice}\label{asec:algPSG}
In this appendix we derive the allowed PSGs for symmetry spin
liquids on the triangular lattice characterized by a $Z_2$ gauge
group. Essentially, we need to make sure that the constraints
imposed by various identities between symmetry group elements are
satisfied by the PSG. In particular the ansatz must satisfy the
relations in equations (\ref{eqn:TT} to \ref{eqn:mirrorrotsix}). It
turns out that if a PSG satisfies all these constraints, then other
relations between symmetry group elements are automatically
satisfied. This is because any group element written as a string
containing products of generators can be brought into a \quo normal
ordered\quo form using just these operations where normal ordering
implies a form of the kind:
$\mirror^{s}[\rotsix]^{r}T_1^{t_1}T_2^{t_2}$ where $s\in \{0,\,1\}$,
$r\in \{0,\,1,\,\ldots5\}$ and $t_1,\,t_2\in {\mathcal Z}$. This is
sufficient to show no new constraints are imposed by other relations
between symmetry operations.

To simplify expressions we introduce forward difference operators
$\dif_1$ and $\dif_2$, defined as $\dif_1 f(r_1,r_2)\equiv
f(r_1+1,r_2)-f(r_1,r_2)$ and $\dif_2 f(r_1,r_2)\equiv
f(r_1,r_2+1)-f(r_1,r_2)$.

We should first consider how PSG changes
if we do a gauge transformation $\PSG{}$ to the ansatz.
After the gauge transformation, the ansatz will be invariant under
$\PSG{}\cdot\PSG{X}X\cdot\PSG{}^{-1}=\PSG{}\cdot\PSG{X}\cdot X\PSG{}^{-1}X^{-1}\cdot X$.
Therefore, $\PSG{X}$ should be replaced by $\PSG{}\PSG{X}\cdot X\PSG{}^{-1}X^{-1}$.
In terms of phases,
\be
\Gphase{X}(\siter ) \rightarrow \Gphase{\PSG{}}(\siter )+\Gphase{X}(\siter )-\Gphase{\PSG{}}[X^{-1}(\siter )]
\label{equ:PSGunderG}
\ee

As in Wen's derivation of fermionic PSG, we make the following assumption:
using gauge freedom, we can make the ansatz explicitly invariant under translation $\Tone$,
or, $\Gphase{\Tone}(\siter)=0$. For triangular lattice, this can be done by solving equations
$\Gphase{\PSG{}}(r_1,r_2)+\Gphase{\Tone}(r_1,r_2)-\Gphase{\PSG{}}(r_1-1,r_2)=0$.
Since $\Gphase{X}$ is a phase, unless explicitly mentioned,
all equations of phases in this section are true modulo $2\pi$.

Then we add the generator $\Ttwo$. Since $\Tone^{-1}\Ttwo\Tone\Ttwo^{-1}=\SGid$,
following the procedure in the example in main text, we have the equation,
\be
\dif_1\Gphase{\Ttwo}(r_1,r_2)=p_1\pi \nn
\ee
where $p_1$ is a site-independent integer.
Due to the $2\pi$ periodicity, $p_1$ has only two distinct choices: 0 and 1,
and $-p_1\equiv p_1\mod 2$.
Other integer parameters, $p_n$ and $p'_n$, in this section also have this property.

Solution of $\Gphase{\Ttwo}$ is then
\be
\Gphase{\Ttwo}(r_1,r_2)=\Gphase{\Ttwo}(0,r_2)+p_1\pi r_1 \nn
\ee

Here we make another assumption:
using the gauge freedom, we can further make $\Gphase{\Ttwo}(0,r_2)=0$,
while preserving $\Gphase{\Tone}(\siter)=0$.
Then the solution simplifies to
\be
\Gphase{\Ttwo}(r_1,r_2)=p_1\pi r_1
\ee

In the above, we made two assumptions,
$\Gphase{\Tone}(\siter)=0$ and $\Gphase{\Ttwo}(0,r_2)=0$.
These two equations can be satisfied if we have open boundary condition.
However great care must be taken for periodic boundary condition before we proceed.
In this section we always assume open boundary condition, and the two assumptions
can be realized by exploiting gauge freedom.

After this gauge fixing procedure
we are still left with three gauge freedoms.
The first one is adding a global constant phase,
\be
\PSG{1}:\quad \Gphase{1}(\siter )=\mathrm{const.}
\ee
According to the equation (\ref{equ:PSGunderG})
this will not change any generator of PSG.
We can use this freedom to fix one of $A_{ij}$ to be real positive.

The second gauge freedom is
\be
\PSG{2}:\quad \Gphase{2}(r_1,r_2)=\pi r_1
\ee
At first sight $\Gphase{\Tone}$ will be changed under this gauge transformation.
According to (\ref{equ:PSGunderG}), after applying $\PSG{2}$,
\begin{eqnarray*}
\Gphase{\Tone}(r_1,r_2)&\rightarrow&\Gphase{2}(r_1,r_2)+\Gphase{\Tone}(r_1,r_2)-\Gphase{2}(r_1-1,r_2)\\
&\rightarrow&\pi
\end{eqnarray*}
However, we are free to add a site-independent constant $\pi$ to any $\Gphase{X}(\siter)$
because of the IGG structure.
So this gauge transformation does not really change $\PSG{\Tone}$ and $\PSG{\Ttwo}$.

Similar to this one, we have the third gauge freedom,
\be
\PSG{3}:\quad \Gphase{3}(r_1,r_2)=\pi r_2
\ee

$\PSG{2}$ and $\PSG{3}$ do not change $\PSG{\Tone}$ and $\PSG{\Ttwo}$,
but will certainly modify other generators of PSG.
Later we will use them to eliminate redundant parameters in our solution.

We can now include generators of point group into consideration.
First consider the reflection: $\mirror$.

Algebraic constraints from $\Tone^{-1}\mirror\Ttwo\mirror^{-1}=\SGid$
and $\Ttwo^{-1}\mirror\Tone\mirror^{-1}=\SGid$ are
\begin{subequations}
\begin{eqnarray*}
\dif_1\Gphase{\mirror}(r_1,r_2)&=&p'_2\pi+p_1 r_2\pi\\
\dif_2\Gphase{\mirror}(r_1,r_2)&=&p'_3\pi+p_1 r_1\pi
\end{eqnarray*}
\end{subequations}
Solution to these equations is
\be
\Gphase{\mirror}(r_1,r_2)=\Gphase{\mirror}(0,0)+p'_2 r_1\pi+p'_3 r_2 \pi+p_1 r_1 r_2 \pi \nn
\ee

Further constraint from $\mirror\mirror=\SGid$ is $p'_2=p'_3\mod 2$ and $2\Gphase{\mirror}(0,0)=p_2\pi$.

Under the gauge transformation $\PSG{2}$, the solution becomes
\begin{eqnarray*}
\Gphase{\mirror}(r_1,r_2)&\rightarrow&\Gphase{2}(r_1,r_2)+\Gphase{\mirror}(r_1,r_2)-\Gphase{2}(r_2,r_1)\\
&\rightarrow&\Gphase{\mirror}(0,0)+(p'_3+1)(r_1+r_2)\pi+p_1 r_1 r_2\pi
\end{eqnarray*}
while $\Gphase{\Tone}$ and $\Gphase{\Ttwo}$ do not change.
Therefore we can always assume that $p'_2=p'_3=0\mod 2$.
Finally we get the general solution (\ref{equ:PSGphimirror}) in the main text.

We are still left with two gauge freedoms at this point.
Both of $\PSG{1}$ and $\PSG{2}\PSG{3}$ do not change
$\PSG{\Tone}$, $\PSG{\Tone}$, and $\PSG{\mirror}$.

Add the last generator $\rotsix$ to the system.
Algebraic constraints from $\Tone^{-1}\rotsix\Ttwo^{-1}\rotsix^{-1}=\SGid$
and $\Ttwo^{-1}\rotsix\Ttwo\Tone\rotsix^{-1}=\SGid$ are
\begin{subequations}
\begin{eqnarray*}
\dif_1\Gphase{\rotsix}(r_1,r_2)&=&p'_4\pi+p_1 r_2\pi\\
\dif_2\Gphase{\rotsix}(r_1,r_2)&=&p'_5\pi+p_1 (r_1-r_2-1) \pi
\end{eqnarray*}
\end{subequations}
Solution to these equations is
\be
\begin{split}
\Gphase{\rotsix}(r_1,r_2)=&\phantom{+}\Gphase{\rotsix}(0,0)+p'_4\pi r_1+p'_5\pi r_2\\
&+p_1 r_1 r_2 \pi+\half p_1 r_2(r_2-1)\pi \nn
\end{split}
\ee

Constraint from $(\rotsix)^{6}=\SGid$ is $12\Gphase{\rotsix}(0,0)=0$.

Further constraint from $\rotsix\mirror\rotsix\mirror=\SGid$ is $p'_4=p'_3=0\mod 2$
and $4\Gphase{\rotsix}(0,0)=0$. Therefore we can assume that $\Gphase{\rotsix}(0,0)=p_3\pi/2$.

Under the gauge transformation $\PSG{2}\PSG{3}$,
the solution of $\Gphase{\rotsix}$ transforms to
\be
\begin{split}
\Gphase{\rotsix}(r_1,r_2)\rightarrow&\phantom{+}\frac{p_3}{2}\pi+(p'_5+1)r_1\pi\\
&+p_1 r_1 r_2\pi+\half p_1 r_2(r_2-1)\pi
\end{split}
\nn
\ee
while $\Gphase{\Tone}$, $\Gphase{\Ttwo}$, and $\Gphase{\mirror}$ do not change.
Therefore, we can always assume that $p'_5=0\mod 2$.
Then we get the general solution (\ref{equ:PSGphirotsix}) in the main text.

A motivation for studying the range of states available
 comes from the numerical simulations\cite{Misguich:ExactDiagTrigLattMultipleSpinPhaseDiagram}
 on the triangular lattice S=1/2 model with ring exchange,
 where a state with a four fold degenerate
 ground state and no magnetic order was found in exact
 diagonalization studies. While this degeneracy is consistent with a
 gapped $Z_2$ spin liquid, the quantum numbers of the degenerate
 states were not compatible with that expected for the simplest such
 spin liquid. The question then arises whether more complicated spin
 liquid states could reproduce the observations. Although we answer this
 question in the negative for the $\pi$-flux state, the nature of a
 phase that could produce the observed degeneracies remains an
 interesting open question.

\section{PSGs of Anisotropic Triangular Lattice}\label{asec:PSGaniso}
Anisotropic triangular lattice can be realized in experimental
systems such as Cs$_2$CuCl$_4$ \cite{Coldea:CsCuCl}. It is also
possible that the spin liquid state breaks triangular lattice space
group symmetry down to anisotropic triangular lattice space group. 
A careful large-N study of Hubbard-Heisenberg model on anisotropic triangular lattice 
has been carried out by Chung, {\it et. al.}\cite{Chung:largeN}. 
They used a unit cell with four sites. Our PSG analysis can actually justify 
their choice of unit cell within symmetric spin liquid states. 

We use the following coordinates system.
\begin{figure}[h]
\includegraphics{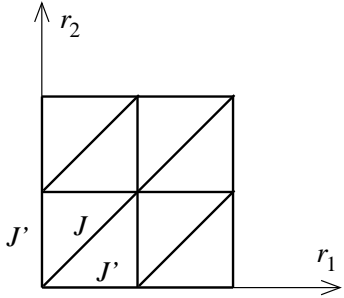}
\caption{\label{fig:anisotrig}
Coordinates system of anisotropic triangular lattice. }
\end{figure}

This space group has also four generators.
The first three are identical to the isotropic case:
$\Tone$, $\Ttwo$, and $\mirror$.
The last one is replaced by a 180 degree rotation $\rottwo$.
\be
\rottwo:\ (r_1,\ r_2)\rightarrow(-r_1,\ -r_2)
\ee

The solution is identical to the isotropic case
before solving the last generator.

For the last generator $\rottwo$,
from constaints $\Tone^{-1}\rottwo\Tone^{-1}\rottwo^{-1}=\SGid$ and
$\Ttwo^{-1}\rottwo\Ttwo^{-1}\rottwo^{-1}=\SGid$, we have two equations
\begin{eqnarray*}
\dif_1\Gphase{\rottwo}(r_1,r_2)=p'_5\pi\\
\dif_2\Gphase{\rottwo}(r_1,r_2)=p_4\pi
\end{eqnarray*}
Solution to these equations is
\be
\Gphase{\rottwo}(r_1,r_2)=\Gphase{\rottwo}(0,0)+p'_5 r_1\pi+p_4 r_2\pi \nn
\ee

Constraint from $\rottwo\rottwo=\SGid$ is $4\Gphase{\rottwo}(0,0)=0$,
then we can assume that $\Gphase{\rottwo}(0,0)=p_3\pi/2$.

Further constraint from $\mirror\rottwo\mirror\rottwo=\SGid$ is $p'_5=p_4\mod 2$.

However we cannot use gauge freedom $\PSG{2}\PSG{3}$ to eliminate $p_4$ this time.

The final solutions are
\begin{subequations}
\begin{eqnarray}
\Gphase{\Tone}(r_1,\ r_2)&=&0\label{equ:PSGanisophiT1}\\
\Gphase{\Ttwo}(r_1,\ r_2)&=&p_1\pi r_1\label{equ:PSGanisophiT2}\\
\Gphase{\mirror}(r_1,\ r_2)&=&p_2\pi/2+p_1 r_1 r_2 \pi\label{equ:PSGanisophimirror}\\
\Gphase{\rottwo}(r_1,\ r_2)&=&p_3\pi/2+p_4\pi(r_1+r_2)\label{equ:PSGanisophirottwo}
\end{eqnarray}
\end{subequations}
where $p_1$, $p_2$, $p_3$, $p_4$ are integers, either 0 or 1.

If nearest-neighbor $A$ are nonzero, we have only two possible realizations.
The difference from isotropic case is, this time not all $A$ are real.
Therefore we can, in principle, have time reversal breaking states.

\begin{figure}[h]
\includegraphics{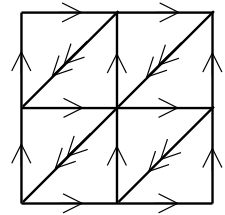}
\caption{\label{fig:anisotrigzeroansatz}
Ansatz of zero-flux state on anisotropic triangular lattice.
Different arrows represent different amplitude of $A$. Single-arrow $A$ are real.
$B$ are not represented in picture.}
\end{figure}

Two realizations:
\quo zero-flux state\quo: $p_1=0,\ p_2=0,\ p_3=1,\ p_4=0$.
\begin{subequations}
\begin{eqnarray*}
\Gphase{\Tone}(r_1,\ r_2)=\Gphase{\Ttwo}(r_1,\ r_2)=\Gphase{\mirror}(r_1,\ r_2)&=&0\\
\Gphase{\rottwo}(r_1,\ r_2)&=&\pi/2
\end{eqnarray*}
\end{subequations}
All $B$ are real, diagonal $A$ can be complex,
$A$ on the other two kinds of bonds are real (see FIG.~\ref{fig:anisotrigzeroansatz}).

\begin{figure}[h]
\includegraphics{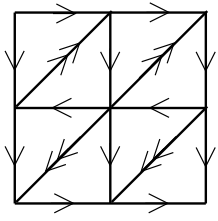}
\caption{\label{fig:anisotrigpiansatz}
Ansatz of $\pi$-flux state on anisotropic triangular lattice.
Different arrows represent different amplitude of $A$. Single-arrow $A$ are real.
$B$ are not represented in picture.}
\end{figure}

\quo $\pi$-flux state\quo: $p_1=1,\ p_2=1,\ p_3=0,\ p_4=1$.
\begin{subequations}
\begin{eqnarray*}
\Gphase{\Tone}(r_1,\ r_2)&=&0\\
\Gphase{\Ttwo}(r_1,\ r_2)&=&\pi r_1\\
\Gphase{\mirror}(r_1,\ r_2)&=&\pi/2+r_1 r_2 \pi\\
\Gphase{\rottwo}(r_1,\ r_2)&=&\pi(r_1+r_2)
\end{eqnarray*}
\end{subequations}
Diagonal $A$ can be complex, diagonal $B$ must be zero.
$A$ on the other two kinds of bonds are real,
$B$ on the other two kinds of bonds are pure imaginary (see FIG.~\ref{fig:anisotrigpiansatz}).

In both states, if diagonal $A$ are neither real nor pure imaginary,
we will have non-trivial flux (not zero or $\pi$) through certain rhombi,
then the ansatz breaks time reversal symmetry.
If we further impose time reversal symmetry,
diagonal $A$ must be real in zero-flux state,
and pure imaginary in $\pi$-flux state.
However, if all $B$ happen to be zero, diagonal $A$ can be
either real or pure imaginary in any of these two states. 

The \quo $\pi$-flux\quo\ symmetric spin liquid state (with $\hat{T}$-symmetry) 
on anisotropic triangular lattice, 
has the same spinon dispersion minima positions with isotropic case, 
therefore they will give rise to the same magnetic ordering wavevectors (M-points: midpoints of B.Z. edges). 

\section{PSGs of Kagom\'e Lattice}\label{asec:PSGKagome}
\subsection{Properties of the different mean field states}
\begin{figure}[h]
\includegraphics{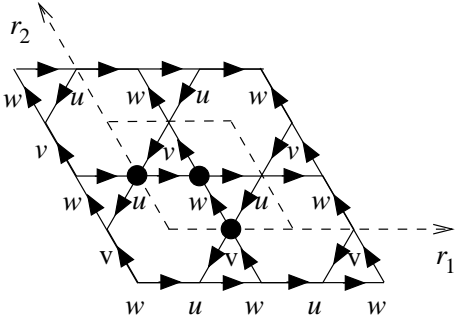}
\caption{\label{fig:Kagome001state} Ansatz of the $p_1=0,\ p_2=0,\
p_3=1$ state which is the $Q_1=-Q_2$ state of Sachdev which on
spinon condensation leads to the $\sqrt{3}\times\sqrt{3}$ state. }
\end{figure}

\begin{figure}[h]
\includegraphics{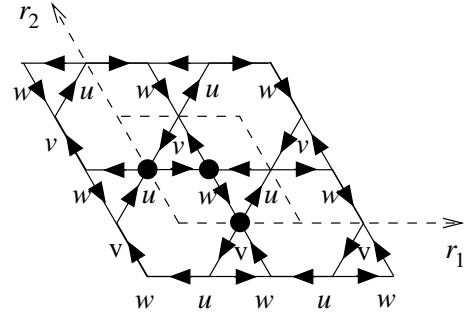}
\caption{\label{fig:Kagome010state} Ansatz of the $p_1=0,\ p_2=1,\
p_3=0$ state which is the $Q_1=Q_2$ state in Sachdev's notation. On
spinon condensation this leads to the $q=0$ magnetically ordered
state. }
\end{figure}

{\em The zero-flux states:} Ansatz for these states are shown in
FIGs \ref{fig:Kagome001state} and \ref{fig:Kagome010state}.

\begin{figure}[h]
\includegraphics{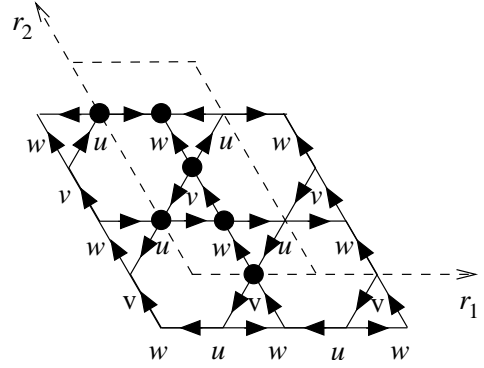}
\caption{\label{fig:Kagome101state}
Ansatz of the $p_1=1,\ p_2=0,\ p_3=1$ state.
Rhombus enclosed by dash lines is the unit cell.
Six sites with black dots form the basis.
}
\end{figure}

{\em 2. $\pi$-Flux States} For the  $p_1=1,\ p_2=0,\ p_3=1$ state,
the ansatz is given in FIG.~\ref{fig:Kagome101state}.

The mean field Hamiltonian for these two $p_1=1$ ($\pi$-flux) states
is similar to that of $\pi$-flux state on triangular lattice,
equations (\ref{equ:FTedmfHpistate}) and (\ref{equ:DPmatrix}). But
the vector spinon field $\spinon$ has twelve components and the
$P(\veck)$ matrix becomes a 6-by-6 hermitian matrix in the following
form \be
P(\veck)=-\im J_1 A_1\begin{pmatrix}\mathcal{A}(\veck)&\mathcal{C}(\veck)\\
\mathcal{C}(\veck)&\mathcal{B}(\veck)
\end{pmatrix}
\nn
\ee
where $\mathcal{A}(\veck),\ \mathcal{B}(\veck),\ \mathcal{C}(\veck)$ are 3-by-3 anti-hermitian matrices.

The $\mathcal{C}(\veck)$ matrix has the same form for both $p_1=1$ states.
\be
\mathcal{C}(\veck)=\begin{pmatrix}
0&e^{\im k_3/2}&-e^{-\im k_2/2}\\
-e^{-\im k_3/2}&0&0\\
e^{-\im k_2/2}&0&0
\end{pmatrix}
\nn
\ee
For $p_1=1,\ p_2=1,\ p_3=0$ state the $\mathcal{A}(\veck)$ matrix is
\be
\mathcal{A}(\veck)=\begin{pmatrix}
0&e^{-\im k_3/2}&-e^{\im k_2/2}\\
-e^{\im k_3/2}&0&2\cos(k_1/2)\\
e^{-\im k_2/2}&-2\cos(k_1/2)&0
\end{pmatrix}
\nn
\ee
The $\mathcal{B}(\veck)$ matrix is
\be
\mathcal{B}(\veck)=\begin{pmatrix}
0&-e^{-\im k_3/2}&-e^{\im k_2/2}\\
e^{\im k_3/2}&0&2\im\sin(k_1/2)\\
e^{-\im k_2/2}&2\im\sin(k_1/2)&0
\end{pmatrix}
\nn
\ee

For $p_1=1,\ p_2=0,\ p_3=1$ state the $\mathcal{A}(\veck)$ matrix is
\be
\mathcal{A}(\veck)=\begin{pmatrix}
0&-e^{-\im k_3/2}&e^{\im k_2/2}\\
e^{\im k_3/2}&0&2\im\sin(k_1/2)\\
-e^{-\im k_2/2}&2\im\sin(k_1/2)&0
\end{pmatrix}
\nn
\ee
The $\mathcal{B}(\veck)$ matrix is
\be
\mathcal{B}(\veck)=\begin{pmatrix}
0&e^{-\im k_3/2}&e^{\im k_2/2}\\
-e^{\im k_3/2}&0&2\cos(k_1/2)\\
-e^{-\im k_2/2}&-2\cos(k_1/2)&0
\end{pmatrix}
\nn
\ee

\subsection{Solution to the Algebraic PSG on the Kagom\'e}
The procedure for solving the algebraic PSG is described below.

The algebraic form of the four space group generators is listed below.
Note the unusual $-1$ in $\rotsix$.
Generators $\Tone$ and $\Ttwo$ preserve sublattices.
\begin{eqnarray*}
\Tone&:\quad&(r_1,r_2)_{p}\rightarrow(r_1+1,r_2)_{p}\\
\Ttwo&:\quad&(r_1,r_2)_{p}\rightarrow(r_1,r_2+1)_{p}
\end{eqnarray*}
$\mirror$ exchanges $\sublatone$ and $\sublattwo$,
\begin{eqnarray*}
\mirror&:\quad&(r_1,r_2)_{\sublatone}\rightarrow(r_2,r_1)_{\sublattwo}\\
&&(r_1,r_2)_{\sublattwo}\rightarrow(r_2,r_1)_{\sublatone}\\
&&(r_1,r_2)_{\sublatthree}\rightarrow(r_2,r_1)_{\sublatthree}
\end{eqnarray*}
$\rotsix$ cyclicly permutates the three sublattices,
\begin{eqnarray*}
\rotsix&:\quad&(r_1,r_2)_{\sublatone}\rightarrow(r_1-r_2-1,r_1)_{\sublattwo}\\
&&(r_1,r_2)_{\sublattwo}\rightarrow(r_1-r_2,r_1)_{\sublatthree}\\
&&(r_1,r_2)_{\sublatthree}\rightarrow(r_1-r_2,r_1)_{\sublatone}
\end{eqnarray*}

As before we can assume that $\Gphase{\Tone,p}(\siter)=0$ and $\Gphase{\Ttwo,p}(0,r_2)=0$, for $p=\sublatone,\sublattwo,\sublatthree$ respectively.

From $\Tone^{-1}\Ttwo\Tone\Ttwo^{-1}=\SGid$, we have the equation
\be
\dif_1\Gphase{\Ttwo,p}(r_1,r_2)=p_1\pi \nn
\ee
for $p=\sublatone,\sublattwo,\sublatthree$ respectively.
Note that the constant $p_1$ must be the same for three sublattices.

Solution is then
\be
\Gphase{\Ttwo,p}(r_1,r_2)=p_1\pi r_1 \nn
\ee

At this stage we are left with more gauge freedom than the triangular case.
The first one is a uniform rotation of boson phases on all sites.
\be
\PSG{1}:\quad \Gphase{1,p}(\siter )=\mathrm{const.} \nn
\ee
As before this guage transformation does not change any element of PSG.
We will use this freedom to fix one of $A_{ij}$({\it e.g.} bond $(0,0)_{\sublatone}-(0,0)_{\sublatthree}$) to be real positive.

The second and third gauge freedoms do not change $\PSG{\Tone}$ and $\PSG{\Ttwo}$.
But they changes the other two generators.
We will use them to fix the form of $\PSG{\mirror}$.
The second one has no correspondence in triangular case,
\be
\begin{split}
\PSG{2}:\quad &\Gphase{2,\sublatone}(\siter )=+\phi_0\\
&\Gphase{2,\sublattwo}(\siter )=-\phi_0\\
&\Gphase{2,\sublatthree}(\siter )=0\\
\end{split}
\nn
\ee
where $\phi_0$ is an arbitrary constant.
The third one is similar to the gauge operation we used in triangular case.
\be
\PSG{3}:\quad \Gphase{3,p}(r_1,r_2 )=\pi r_1
\nn
\ee
where $p=\sublatone,\sublattwo,\sublatthree$.

The fourth gauge freedom does not change $\PSG{\Tone}$, $\PSG{\Ttwo}$ and $\PSG{\mirror}$.
But it changes $\PSG{\rotsix}$. We will use this freedom to fix the form of $\PSG{\rotsix}$.
\be
\begin{split}
\PSG{4}:\quad &\Gphase{4,\sublatone}(r_1,r_2 )=\pi (r_1+r_2)\\
&\Gphase{4,\sublattwo}(r_1,r_2 )=\pi (r_1+r_2)\\
&\Gphase{4,\sublatthree}(r_1,r_2 )=\pi (r_1+r_2+1)
\end{split}
\nn
\ee

We can now introduce point group generator $\mirror$.
Algebraic constraints from $\Tone^{-1}\mirror\Ttwo\mirror^{-1}=\SGid$
and $\Ttwo^{-1}\mirror\Tone\mirror^{-1}=\SGid$ are
\begin{subequations}
\begin{eqnarray*}
\dif_1\Gphase{\mirror,p}(r_1,r_2)&=&p'_2\pi+p_1 r_2\pi\\
\dif_2\Gphase{\mirror,p}(r_1,r_2)&=&p'_3\pi+p_1 r_1\pi
\end{eqnarray*}
\end{subequations}
where $p=\sublatone,\sublattwo,\sublatthree$, $p'_2$ and $p'_3$ are integer constant independent
of unit cell index $(r_1,r_2)$ and sublattice index $p$.
Solution to these equations is
\be
\Gphase{\mirror,p}(r_1,r_2)=\Gphase{\mirror,p}(0,0)+p'_2 r_1\pi+p'_3 r_2 \pi+p_1 r_1 r_2 \pi \nn
\ee
However, we have no {\it a priori} reason to say $\Gphase{\mirror,p}(0,0)$ are independent of sublattices index $p$.

Further constraint from $\mirror\mirror=\SGid$ is $p'_2=p'_3\mod 2$ and
\begin{eqnarray*}
2\Gphase{\mirror,\sublatthree}(0,0)&=&p_2\pi\\
\Gphase{\mirror,\sublatone}(0,0)+\Gphase{\mirror,\sublattwo}(0,0)&=&p_2\pi
\end{eqnarray*}
where $p_2$ is an integer constant.
This fixes $\Gphase{\mirror,\sublatthree}(0,0)$ to be $p_2\pi/2$ but leaves one freedom for $\Gphase{\mirror,\sublatone}(0,0)$ and $\Gphase{\mirror,\sublattwo}(0,0)$. As in triangular case we can use $\PSG{3}$ to make $p'_2=p'_3=0$. Because after applying $\PSG{3}$ the solution to $\Gphase{\mirror}$ becomes, according to (\ref{equ:PSGunderG}),
\be
\Gphase{\mirror,p}(r_1,r_2)\rightarrow\Gphase{\mirror,p}(0,0)+(p'_2+1) (r_1+r_2)\pi+p_1 r_1 r_2\pi \nn
\nn
\ee
We now apply $\PSG{2}$.
\begin{eqnarray*}
\Gphase{\mirror,\sublatone}(r_1,r_2)&\rightarrow&\Gphase{\mirror,\sublatone}(0,0)+2\phi_0+p_1 r_1 r_2\pi\\
\Gphase{\mirror,\sublattwo}(r_1,r_2)&\rightarrow&\Gphase{\mirror,\sublattwo}(0,0)-2\phi_0+p_1 r_1 r_2\pi\\
\Gphase{\mirror,\sublatthree}(r_1,r_2)&\rightarrow&\Gphase{\mirror,\sublatthree}(0,0)+p_1 r_1 r_2\pi
\end{eqnarray*}
Therefore we can always make $\Gphase{\mirror,\sublatone}=\Gphase{\mirror,\sublattwo}=p_2\pi/2$ by choosing appropiate $\phi_0$. Then the three phase functions $\Gphase{\mirror,p}$ are identical.

Add the last generator $\rotsix$ to the system.
Algebraic constraints from $\Tone^{-1}\rotsix\Ttwo^{-1}\rotsix^{-1}=\SGid$
and $\Ttwo^{-1}\rotsix\Ttwo\Tone\rotsix^{-1}=\SGid$ are
\begin{subequations}
\begin{eqnarray*}
\dif_1\Gphase{\rotsix,p}(r_1,r_2)&=&p'_4\pi+p_1 r_2\pi\\
\dif_2\Gphase{\rotsix,p}(r_1,r_2)&=&p'_5\pi+p_1 (r_1-r_2-1) \pi
\end{eqnarray*}
\end{subequations}
Solution to these equations is
\be
\begin{split}
\Gphase{\rotsix,p}(r_1,r_2)=&\phantom{+}\Gphase{\rotsix,p}(0,0)+p'_4\pi r_1+p'_5\pi r_2\\
&+p_1 r_1 r_2 \pi+\half p_1 r_2(r_2-1)\pi \nn
\end{split}
\ee
Again we should not assume that $\Gphase{\rotsix,p}(0,0)$ are independent of sublattices index $p$.

Further constraint from $\rotsix\mirror\rotsix\mirror=\SGid$ is $p'_4=p'_3=0\mod 2$ and
\begin{eqnarray*}
2\Gphase{\rotsix,\sublattwo}(0,0)&=&p_3\pi\\
\Gphase{\rotsix,\sublatone}(0,0)+\Gphase{\rotsix,\sublatthree}&=&p_3\pi
\end{eqnarray*}
Therefore we have $\Gphase{\rotsix,\sublattwo}(0,0)=p_3\pi/2$.

A similar constraint from $\mirror\rotsix\mirror\rotsix=\SGid$ is $p'_4=p'_3=0\mod 2$ and
\begin{eqnarray*}
2\Gphase{\rotsix,\sublatone}(0,0)&=&p_4\pi\\
\Gphase{\rotsix,\sublattwo}(0,0)+\Gphase{\rotsix,\sublatthree}&=&p_4\pi
\end{eqnarray*}
Combining with previous constraints on $\Gphase{\rotsix,p}$ we conclude that
$p_4\equiv p_3\mod 2$ and $\Gphase{\rotsix,p}=p_3\pi/2$.

Under the gauge transformation $\PSG{4}$,
the solution of $\Gphase{\rotsix}$ transforms to
\be
\begin{split}
\Gphase{\rotsix,p}(r_1,r_2)\rightarrow&\phantom{+}\frac{p_3}{2}\pi+\pi+(p'_5+1)r_1\pi\\
&+p_1 r_1 r_2\pi+\half p_1 r_2(r_2-1)\pi
\end{split}
\nn
\ee
while $\Gphase{\Tone,p}$, $\Gphase{\Ttwo,p}$, and $\Gphase{\mirror,p}$ do not change.
Therefore, we can always assume that $p'_5=0\mod 2$.
The resulting constant $\pi$ can be neglected because it is an IGG operation.
Then we get the general solution (\ref{equ:PSGKagomephirotsix}).


\begin{thebibliography}{}
\bibitem{incommensurate}{For larger
values of next neighbor coupling $\alpha>0.31$, the location of the
spinon minima jump to an incommensurate value, which we do not
discuss here.}
\bibitem{Spin} {The expression for the permutation operator on four spin-1/2 can be written as:
$P_4+h.c.=4[
 (\mathbf{S}_i\cdot\mathbf{S}_j)(\mathbf{S}_k\cdot\mathbf{S}_l)
-(\mathbf{S}_i\cdot\mathbf{S}_k)(\mathbf{S}_j\cdot\mathbf{S}_l)
+(\mathbf{S}_i\cdot\mathbf{S}_l)(\mathbf{S}_j\cdot\mathbf{S}_k)
 ]
+(1/2)\left (\mathbf{S}_i+\mathbf{S}_j+\mathbf{S}_k+\mathbf{S}_l\right )^2-(5/4)$
}
\end{thebibliography}
\end{document}